\documentclass[aps,pre,twocolumn,10pt,superscriptaddress,showpacs,floatfix]{revtex4-1}

\usepackage{amssymb}
\usepackage{epsfig}
\usepackage{multirow}
\usepackage{graphicx}
\usepackage[table]{xcolor}

\begin{document}
\title{Sampling properties of directed networks}

\author{S.-W. Son} \thanks{Corresponding author: sonswoo@hanyang.ac.kr}
\affiliation{Complexity Science Group, University of Calgary, Calgary T2N 1N4, Canada}
\affiliation{Department of Applied Physics, Hanyang University, Ansan 426–791, Korea}
\author{C. Christensen} \affiliation{Complexity Science Group, University of Calgary, Calgary T2N 1N4, Canada}
\author{G. Bizhani} \affiliation{Complexity Science Group, University of Calgary, Calgary T2N 1N4, Canada}
\author{D. V. Foster} \affiliation{Complexity Science Group, University of Calgary, Calgary T2N 1N4, Canada}
\author{P. Grassberger} \affiliation{Complexity Science Group, University of Calgary, Calgary T2N 1N4, Canada} \author{M. Paczuski} \affiliation{Complexity Science Group, University of Calgary, Calgary T2N 1N4, Canada}

\date{\today}

\begin{abstract}
 For many real-world networks only a small ``sampled" version of
the original network may be investigated; those results are then
used to draw conclusions about the actual system.  Variants of
breadth-first search (BFS) sampling, which are based on epidemic
processes, are widely used. Although it is well established that
BFS sampling fails, in most cases, to capture the IN-component(s)
of directed networks, a description of the effects of BFS sampling
on other topological properties are all but absent from the
literature. To systematically study the effects of sampling biases
on directed networks, we compare BFS sampling to random sampling
on complete large-scale directed networks.  We present new results
and a thorough analysis of the topological properties of seven
different complete directed networks (prior to sampling),
including three versions of Wikipedia, three different sources of
sampled World Wide Web data, and an Internet-based social network.
We detail the differences that sampling method and coverage can
make to the structural properties of sampled versions of these
seven networks. Most notably, we find that sampling method and
coverage affect both the bow-tie structure, as well as the number
and structure of strongly connected components in sampled
networks. In addition, at low sampling coverage ({\em i.e.} less
than 40\%), the values of average degree, variance of out-degree,
degree auto-correlation, and link reciprocity are overestimated by
30\% or more in BFS-sampled networks, and only attain values
within 10\% of the corresponding values in the complete networks
when sampling coverage is in excess of 65\%. These results may
cause us to rethink what we know about the structure, function,
and evolution of real-world directed networks.
\end{abstract}
\maketitle

\section{Introduction}

In the last decade, a flood of research on systems that can be
represented as networks has revealed that most differ markedly
from simple random graph
models~\cite{Strogatz2001,Albert2002,Dorogovtsev2002,Newman2003}.
For example, many exhibit a broad, or ``scale-free" degree
distribution, making them robust to random failures but rendering
them vulnerable to targeted
attacks~\cite{Albert2000,Cohen2000,Cohen2001}. Complex networks
research also offers a framework for representing biological
processes such as gene regulation~\cite{Davidson2002},
protein-protein interactions~\cite{HJeong2001,Mering2002}, and
even connections between diseases and
symptoms~\cite{KGoh2007,Yildirim2007}. Our growing understanding
of how epidemics spread on networks has led to parallel insights
into the propagation of information, fashions, ideas, and
fads~\cite{Grassberger1983,PastorSatorras2001,PastorSatorras2001a,Zanette2001,Kuperman2001,Moreno2004,Lind2007}.
Complex network studies have incorporated features such as the
weight and direction of links to describe systems more precisely.
Link directionality plays a particularly important role in
dynamics, as small changes to link structure can completely change
the dynamics on a
network~\cite{DUHwang2005,Nishikawa2006,SMPark2006,SWSon2009}.
Thus, capturing the directional structure is essential to
understanding the dynamics on directed networks, much as the
connectivity structure is essential to dynamic processes on
undirected networks.

 One impediment, however, is that it is difficult, if
not impossible, to obtain a complete list of links for many
networks, including, for example, the World Wide Web (WWW) or
large-scale gene-regulatory networks. The Web changes so quickly
that by the time one could have covered it, it would be
substantially transformed~\cite{Ntoulas2004,WWWSize}.  Even if one
somehow managed to completely map its structure at some point in
time, analyzing such a large network, estimated to contain at
least 19 billion pages~\cite{WWWSize}, would present further
impediments. There is no way to avoid biases because the Web can
only be sampled by following directional hyperlinks, which leaves
portions inaccessible.  Furthermore, as has been found to be the
case with sampled, undirected networks, the sampled Web's
appearance might even fundamentally change depending on the type
of sampling method used~\cite{SamplingBook,Newman2003a}. If we are
to have a clear and reliable picture of large-scale directed
networks and their statistical properties,  it is important that
we quantitatively understand the effects of sampling biases on the
properties of interest, as well as why such biases arise. Insight
into these questions stands to impact structure-exploiting search
and ranking algorithms, such as Google's PageRank~\cite{PageRank,Son2012},
and may cause us to rethink what we know about the structure,
function, and evolution of real-world networks.

Up to now the statistical properties of sampled {\em undirected
networks} have been investigated in several papers. Stumpf {\it et
al.}~\cite{Stumpf2005,Stumpf2005a} studied the degree distribution
of two random networks -- one that had been sampled ``uniformly"
by picking nodes at random, and one that was subject to
connectivity-dependent sampling -- both analytically and
numerically. Lee {\it et al.}~\cite{SHLee2006} also studied,
numerically, the effects of random sampling and snowball
sampling~\cite{snowball}, on statistical properties of real
scale-free networks -- including degree distribution exponents,
betweenness centrality exponents, assortativity, and clustering
coefficient -- demonstrating that these quantities could be either
overestimated or underestimated, depending on the fraction of the
network sampled and the type of sampling method used. Kurant {\it
et al.}~\cite{Kurant2010,Kurant2011} provided a detailed analytic
treatment of measurement bias due to random sampling and
breadth-first search (BFS) sampling~\cite{BFS}, showing that, for
example, BFS sampling can lead to overestimation of average
degree, and random sampling to underestimation.

In the case of directed networks, however, even though the
structural properties of many directed real-world networks have
been examined~\cite{Dorogovtsev2001,Garlaschelli2004,Bianconi2008,
Leicht2008,Donato2008}, much less work has addressed the
statistical properties that result from sampling
them~\cite{Becchetti2006,Wangt2010}. The majority of numerical
studies in these works were not performed at a level of rigor
sufficient to produce meaningful statistics. Also, the researchers
began their studies with already-biased network data since they
were obtained using a web crawler.

For these reasons, we investigate the biases induced by sampling
large-scale directed networks starting with complete networks that
differ from one another structurally. Several structural
properties such as average degree, variance in degree, degree
auto-correlation, reciprocity, assortativity, and component
structure -- all of which are defined in later sections -- are
analyzed to give a more complete picture of sampling-induced
biases. Because we know the full, final structure we can
accurately measure how systematic errors in measured quantities
are affected by sampling coverage and sampling method. We find
that the earlier conclusions in~\cite{SHLee2006,Kurant2011}
regarding biases in average degree of undirected networks due to
random sampling and BFS sampling also hold for directed networks.
On the other hand, in direct contradiction
with~\cite{Becchetti2006}, we conclude that both random sampling
and BFS sampling overestimate edge reciprocity in the networks we
study. We show that both sampling methods overestimate degree
auto-correlation, sometimes by nearly 400\%. In addition, we find
that random sampling and BFS sampling affect the variance of in-
and out-degree differently: both are underestimated by random
sampling while variance in in-degree is underestimated and
variance in out-degree overestimated by BFS sampling. Finally, we
expand on the work in~\cite{Becchetti2006} by providing a thorough
examination of component sizes and abundances under random
sampling and BFS sampling.

In Sec.~\ref{Properties_of_Directed Networks}, we define the large-scale structural properties of
directed networks, the so-called ``bow-tie"
structure~\cite{Broder2000}, and introduce the complete networks
studied in this work. We also provide a detailed accounting of the
sampling methods used. In Sec.~\ref{Sampling_Results}, we present results for BFS
sampling and uniform, random sampling on these networks and, where
possible, provide arguments regarding how sampling can lead to
measurement bias. We systematically study how accuracy is affected
by the fraction of the network sampled in the two cases. Finally,
summary and concluding remarks are given in Sec.~\ref{Summary_and_Discussion}.

\section{Properties of Directed Networks}
\label{Properties_of_Directed Networks}

\subsection{Directed Networks and Bow-Tie Structure}

\begin{figure}[b]
\centering
\includegraphics[width=0.48\textwidth]{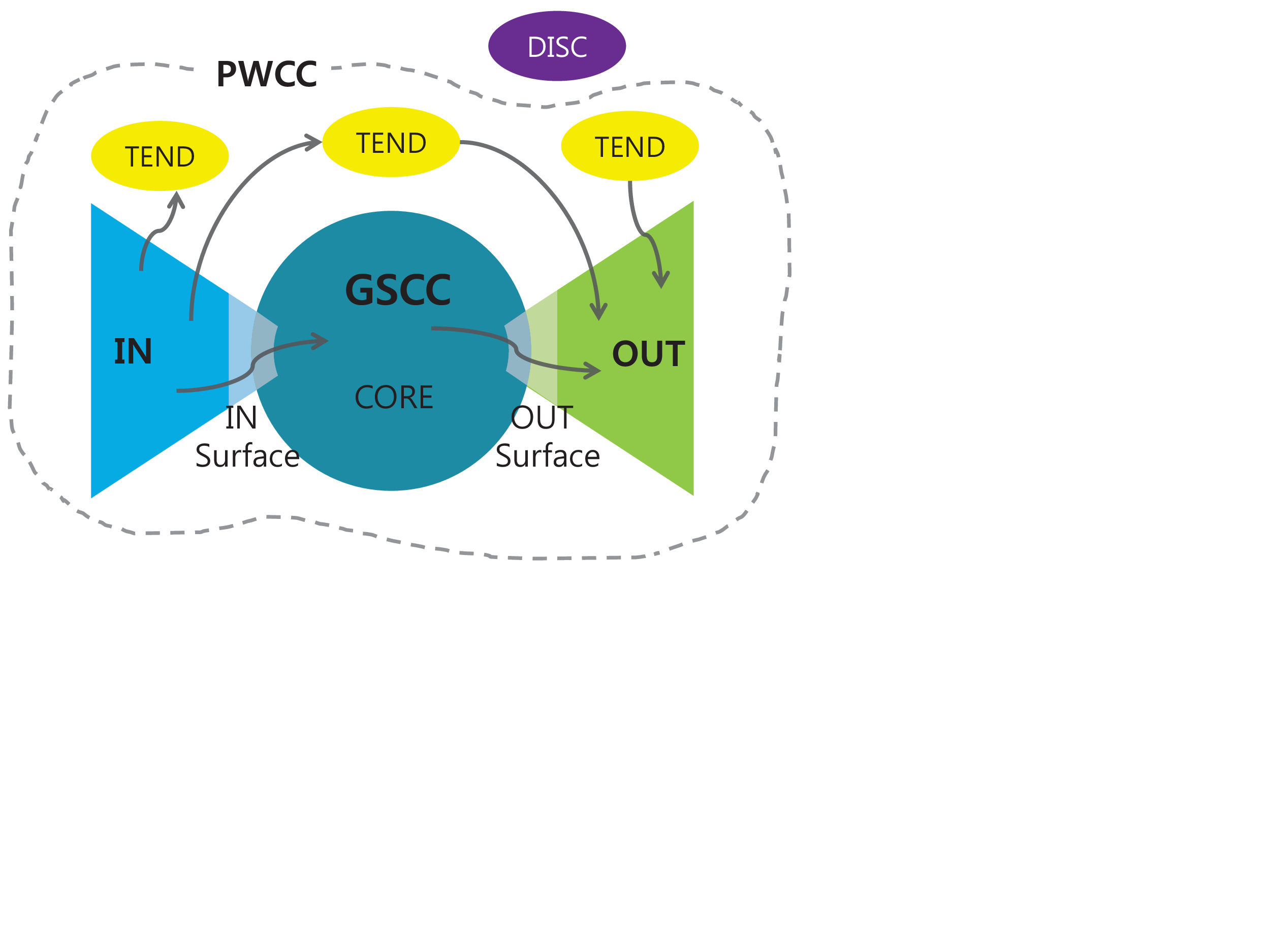}
\caption{ (Color online) Schematic of components and surfaces of a
directed network.  The giant strongly connected component (GSCC)
and the in- and out-components (IN and OUT) account for the
``bow-tie" structure.  Together with tendrils (TEND), these
components form the primary weakly connected component (PWCC).
Portions of the network that are not connected to the PWCC are
disconnected components (DISC).  Nodes of the GSCC that are
directly connected to nodes of IN or OUT form the surface of the
GSCC;  the nodes in IN or OUT to which GSCC surface nodes connect
form the IN and OUT surfaces.} \label{fig:compmap}
\end{figure}

\begin{table*}[th!]%
\begin{minipage}{0.8\textwidth}
\caption{\label{summary1} Summary of component ratios for the data
sets. $N_0$ is the total number of nodes in the networks,
$N_{SCC}$ is the number of SCCs in the networks, and the component
ratios mean how many nodes are placed in each component. Surface
nodes ratio is the percentage of the surface nodes.}
\centering%
\rowcolors{3}{gray!35}{}
\renewcommand{\tabcolsep}{0cm}
\renewcommand{\arraystretch}{1.2}
\begin{tabular}{rrrrrrrrr}
\hline
\multirow{2}*{~Network~~} & \multirow{2}*{~~~$N_0$~~~} & \multirow{2}*{~~~$N_{\rm SCC}$~} &  \multicolumn{4}{c}{\footnotesize Components (\%)} &  \multicolumn{2}{c}{\footnotesize Surface Nodes (\%)} \\
 & & & \scriptsize~~~~GSCC & \scriptsize~~OUT & \scriptsize~~~~~IN~ & \scriptsize~TEND & \scriptsize~~~in GSCC & \scriptsize~~in PWCC~ \\
\hline
~BerkStan~~ & 654,782~ & 107,858 & 51.1 & 19.1 & 24.4 & 5.4 & 9.6 & 20.0~ \\
~Google~~ & 855,802~ & 355,451 & 50.8 & 19.4 & 21.1 & 8.7 & 41.9 & 52.2~ \\
~Stanford~~ & 255,265~ & 29,157 & 59.0 & 26.4 & 12.8 & 1.7 & 7.4 & 18.3~ \\
~RaySoda~~ & 17,852~ & 10,455 & 39.6 & 14.1 & 39.6 & 6.7 & 75.3 & 80.8~ \\
~Wiki2005~~ & 1,596,970~ & 490,715 &69.1& 2.2 & 28.6 & 0.1 & 43.5 & 59.8~ \\
~Wiki2006~~ & 2,935,761~ & 928,028 & 68.3 & 1.5 & 30.1 & 0.2 & 45.1 & 61.1~ \\
~Wiki2007~~ & 3,512,462~ & ~1,148,923 & 67.2 & 1.4 & 31.4 & 0.1 & 46.5 & 62.4~ \\
\hline
\end{tabular}
\end{minipage}
\end{table*}

If one explores a directed network by following links, some
portions of the network are reachable while other portions may not
be. It might be possible to go from one site to another, while the
return journey is impossible. This results in a component picture
of a directed network, as shown in Fig.~\ref{fig:compmap}. We can
define a set of nodes among which a path both to and from all
other nodes in the set exists. This is a  {\em strongly connected
component} (SCC)~\cite{Tarjan1972}.  A directed network can be
decomposed into SCCs if isolated nodes, or nodes with only a
single incoming or outgoing link are considered to be their own
SCC. Then  {\em Tarjan}'s algorithm~\cite{Tarjan1972} can be
easily modified to identify the network's SCCs. The largest SCC is
called the {\em giant strongly connected component} (GSCC), and
corresponds to the knot in the so-called {\em bow-tie
structure}~\cite{Broder2000}. However, we can also ignore link
directionality and identify the sets of nodes that are connected.
These are {\em weakly connected components} (WCC). A fragmented
network may contain several WCCs; the largest of these is called
the {\em giant weakly connected component}
(GWCC)~\cite{Dorogovtsev2001} and the WCC which contains the GSCC
is defined as the {\em primary weakly connected component} (PWCC).
Usually the PWCC is identical to the GWCC.

The {\em out-component(s)} (OUT) of a network are found by
starting from the GSCC and following outgoing links. All those
nodes that can be reached from the GSCC but that do not have paths
back are part of an out-component. Conversely, all nodes that can
reach the GSCC following directed links, but that cannot be
reached from it form the {\em in-component(s)} (IN) of the
network. The IN and OUT correspond to the two wings of the
bow-tie, shown in Fig.~\ref{fig:compmap}.  All other nodes that
are in the PWCC but that are not themselves part of the GSCC, IN,
or OUT form {\em tendrils} (TEND). (Note that our definition of
TEND is not the same as in previous
works~\cite{Becchetti2006,Broder2000}, as we include within
tendrils what they call {\it tubes} -- direct bridges between IN
and OUT.) Any other nodes in the network must be disconnected from
the PWCC and are therefore said to be {\em disconnected
components} (DISC). The GSCC connects with IN and OUT through {\em
surfaces} of these components.  The GSCC-surface is comprised of
the nodes in the GSCC that share links with nodes in IN or OUT
components; nodes in IN that adjoin the GSCC form the IN-surface;
and the nodes in OUT that abut the GSCC form the OUT-surface. The
set of nodes in the GSCC, excluding the surface nodes, is its {\em
core}~\cite{Donato2008,Levene2004} (see Fig.~\ref{fig:compmap}).
Cores for IN and OUT can also be defined. Broder {\it et al.}
reported that $~30\%$ of their sampling of the WWW is GSCC, while
IN and OUT each have roughly 23\% of the nodes~\cite{Broder2000}.
This type of result varies strongly from network to network. Such
differences between many real-world, directed networks are pointed
out in the next section.

\subsection{Data Sets}

\begin{figure*}[th!]
\centering
\includegraphics[width=0.85\textwidth]{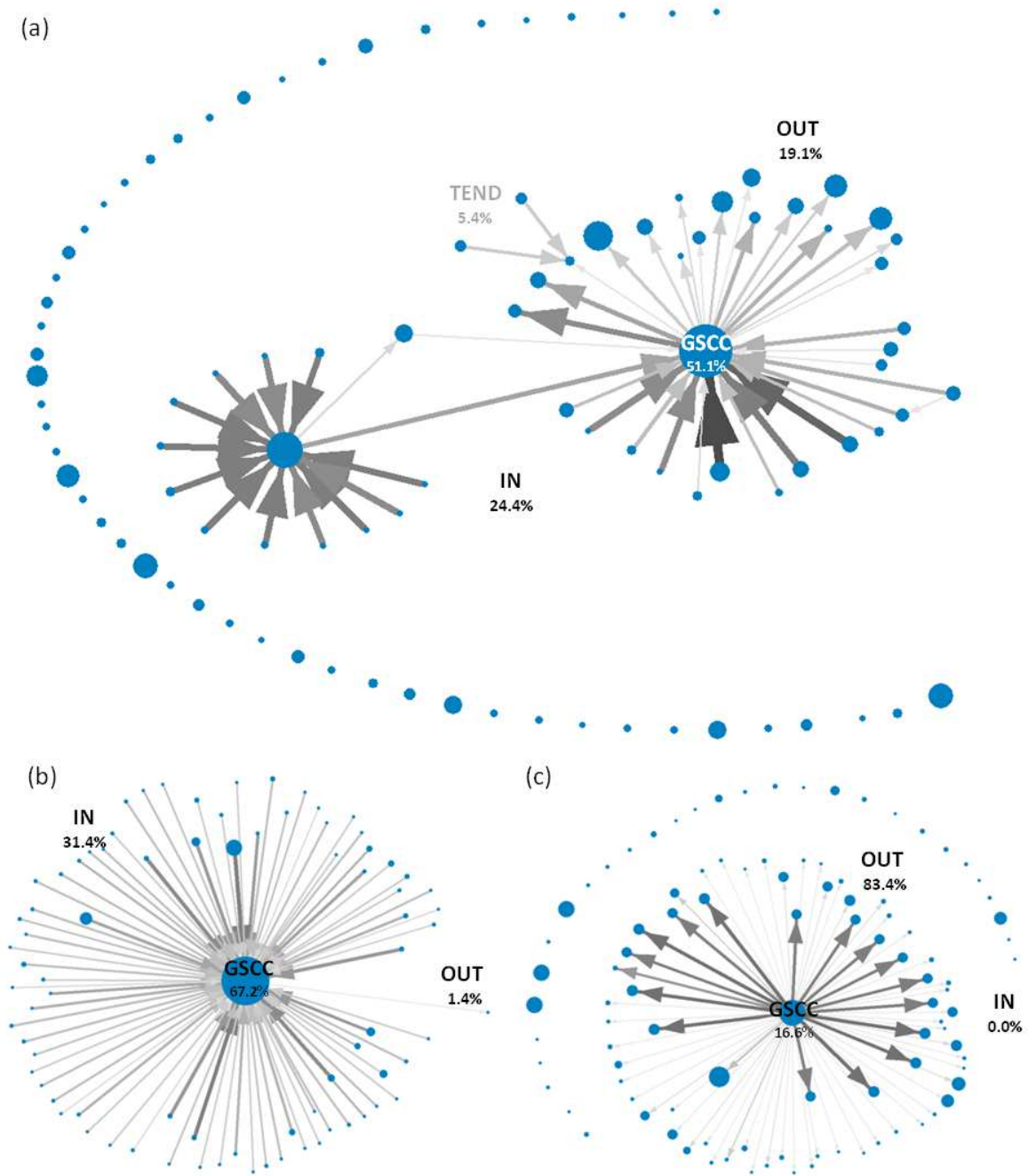}
\caption{ (Color online) SCC diagram for (a) BerkStan
data~\cite{StanfordDataset,Leskovec2008}, (b)
Wiki2007~\cite{WikiDataset}, and (c) Notre Dame
data~\cite{Albert1999b} (not otherwise analyzed, but shown here
for its distinctive structure). For better visualization, only the
100 largest SCCs have been displayed. Each circle corresponds to a
SCC, whose size is proportional to the logarithm of the number of
nodes in the SCC. The width and intensity of the color of the
directed links are proportional to their weight, and self-links
are omitted. This SCC diagram shows the heterogeneity that can
exist for the simple  bow-tie diagram. } \label{fig:BerkStan_SCC}
\end{figure*}

\begin{table*}[th!]
\begin{minipage}{0.8\textwidth}
\caption{\label{summary2} Summary of the basic network properties
for the data sets. $N_0$ is the number of nodes in the GWCC;
$\langle k_{\rm in(out)} \rangle$ is the average incoming
(outgoing) degree, always $\langle k_{\rm in} \rangle = \langle k_{\rm out} \rangle$; $\sigma^2_{\rm in}$ and $\sigma^2_{\rm out}$
are the variances of in- and out-degree, respectively; $r_{a}$ is
the degree auto-correlation; $R$ is the global reciprocity; and
$r_{\rm ii}$, $r_{\rm io}$, $r_{\rm oi}$, and $r_{\rm oo}$ are the
in-degree/in-degree, in-degree/out-degree, out-degree/in-degree,
and out-degree/out-degree assortativities.}
\centering%
\rowcolors{2}{}{gray!35}
\renewcommand{\tabcolsep}{0cm}
\renewcommand{\arraystretch}{1.2}
\begin{tabular}{@{}*{12}{r}}
\hline
~Network~ & $N_0$~~~ & \footnotesize~~$\langle k_{\rm in} \rangle$ & $\sigma^2_{\rm in}$ & $\sigma^2_{\rm out}$ & $r_{a}$ & $R$ & $r_{\rm ii}$ & $r_{\rm io}$ & $r_{\rm oi}$ & $r_{\rm oo}$~ \\
\hline
~BerkStan~ & 654,782 & ~~~11.45 & ~~~84,871.8 & 276.3 & ~~~0.043 & ~~~0.244
& ~~~-0.011 & ~~0.036 & ~~-0.184 & ~~~0.481~ \\
~Google~ & 855,802 & ~5.92 & 1,573.0 & 43.9 & 0.136 & 0.306 &
-0.014 & 0.033 & -0.066 & 0.056~ \\
~Stanford~ & 255,265 & 8.75 & 30,532.2 & 137.6 & 0.046 & 0.262 &
-0.013 & 0.007 & -0.134 & 0.031~ \\
~RaySoda~ & 17,852 & 9.47 & 2,995.3 & 268.2 & 0.331 & 0.202 &
-0.048 & 0.048 & -0.125 & 0.093~ \\
~Wiki2005~ & ~1,596,970 & 12.37 & 43,931.4 &
985.1 & 0.203 & 0.122 & -0.014 & 0.017 & -0.070 & -0.032~ \\
~Wiki2006~ & ~2,935,761 & 12.69 & 56,821.4 &
1,095.8 & 0.196 & 0.118 & -0.008 & 0.014 & -0.051 & -0.034~ \\
~Wiki2007~ & ~3,512,462 & 12.82 & ~~~63,526.7 &
~~~1,101.7 & 0.198 & 0.118 & -0.007 & 0.013 & -0.048 & -0.032~ \\
\hline
\end{tabular}
\end{minipage}
\end{table*}

We analyze seven networks: three sampled Web data sets from
different sources, one complete social network, and three versions
of the entire English language Wikipedia network. The Web data is
a combined set of Web pages from the University of California
Berkeley (berkeley.edu) and Stanford University (stanford.edu),
denoted by ``BerkStan'', Web pages solely from Stanford University
(stanford.edu), denoted by ``Stanford'', and a set of Web pages
released by Google in 2002 as a part of the Google Programming
Contest. All three of these data sets are available for download
from the Stanford Large Network Dataset
Collection~\cite{StanfordDataset,Leskovec2008}. In addition, we
have gathered social network data from an amateur photographers'
website, RaySoda~\cite{RaySoda}, where each node corresponds to a
photographer, and where a directed link from A to B indicates that
A follows B. The largest networks we analyze are the Wikipedia
networks~\cite{Wikipedia} ($\sim \mathcal{O}(10^6) $ nodes) --
three networks collected at different times (2005, 2006, and
2007). These networks, downloaded from~\cite{WikiDataset}, contain
nodes representing five types of Wikipedia page:  articles,
categories, portals, disambiguations, and redirects~\cite{note}.
The number of nodes in our networks is different from those
in~\cite{Buriol2006} since the networks in~\cite{Buriol2006}
contain only article pages, while ours contain the full collection
of pages in the ``main" name space of Wikipedia.

We have elected these seven networks for analysis, not only
because they vary substantially in size, but also because they
have different structural properties. As can be seen in
Fig.~\ref{fig:BerkStan_SCC} and Table ~\ref{summary1}, the
relative sizes of components can span a wide range:  the BerkStan
data epitomize the classical bow-tie, with the bulk of nodes
residing in the GSCC and the remainder balanced between IN and
OUT; the Wikipedia networks, on the other hand, display almost no
OUT, but instead show a tendency for roughly 67\% of the nodes to
comprise the GSCC, and the rest, the IN; conversely, the nodes of
the Notre Dame data~\cite{Albert1999b}   (which is shown in
Fig.~\ref{fig:BerkStan_SCC}(c) for comparison, but is otherwise
not analyzed in this paper), depicting webpages within the nd.edu
domain tend to concentrate in the OUT, revealing no IN and a GSCC
containing less than 20\% of the network's mass. This structure
reflects how that dataset was obtained:  webpages were gathered by
crawling outward from a particular starting page.

Remarkably, even with these strong differences in gross global
structure, we find, as shown in the next sections, many common
trends in the effects of sampling biases on the measured
properties of these networks.  For all data sets,  basic network
properties, including degree distributions, average degree,
variance in in- and out-degrees, degree auto-correlation,
reciprocity, and four types of assortativity, are determined, and
these properties, as well as component analyses, are defined and
presented in corresponding subsections on sampling.  All basic
properties are summarized in Tables \ref{summary1} and
\ref{summary2}, and the values reported therein are later used for
comparison with our sampling studies.  Because it will be
necessary  to avoid trivial sampling failures (resulting from, for
example, network disconnectedness), we consider for analysis only
the GWCC of each network, which by virtue of the fact that, in all
cases, it contains more than 90\% of the network, is also the
PWCC.

\subsection{Sampling Methods}

\begin{figure}[b]
\includegraphics[width=0.45\textwidth]{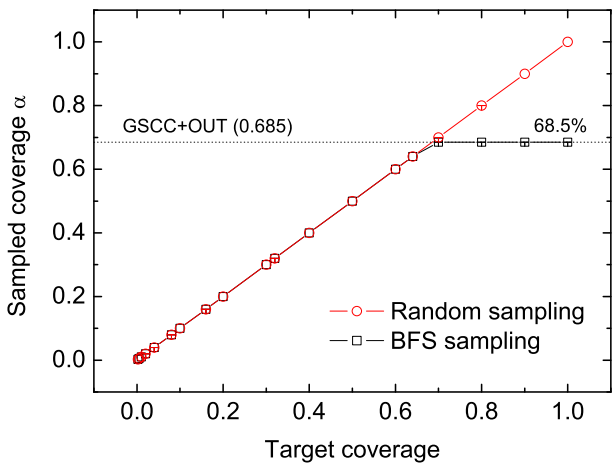}
\caption{ (Color online) Sampled coverages using two different
sampling methods. While random sampling is able to achieve the
targeted coverage, in most cases, BFS only covers up to the size
of the GSCC and OUT. The graph shows the 2007 Wikipedia data which
contains 3,512,462 nodes, and for which the combined fraction of
nodes in GSCC and OUT is about 68.5\%. } \label{fig:method}
\end{figure}

We use two sampling methods: {\em uniform random sampling} and
{\em breadth-first search} (BFS) {\em sampling}. For the former,
each node is selected independently and with equal probability.
This method is not feasible on the real WWW, but it is a good
basis for comparison since it is analytically tractable and is
related to well-known percolation
phenomena~\cite{Albert2000,Cohen2000,Cohen2001}. The latter method
is more complex but has been broadly adopted for web crawling, and
so is important to
analyze~\cite{SamplingBook,Newman2003a,Kurant2010,Kurant2011}.
Starting from a few randomly-selected nodes ({\em seeds}),
neighboring nodes connected by outgoing links are visited at each
successive step like a process of gossip spreading~\cite{Lind2007}. At the outset, the seeds are added to the BFS
queue. One at a time, the outgoing links of these seeds are
explored, and the visited neighbours are added to the queue. We
define the {\em growing front} nodes to be those nodes in the
sampled network whose outgoing links have not yet been explored --
{\it i.e.} those nodes most recently added to the queue. Before
sampling begins, a targeted sampling coverage -- the fraction of
the network one wishes to sample -- is also chosen.  When this
coverage is reached, the process terminates and all edges
connecting already visited nodes are included as part of the final
sampled network. This procedure is analogous to web crawling,
initiating with several portal pages from which Web pages are
iteratively gathered.

While BFS sampling will always cover the entire network in an
undirected (connected)  network, this is not the case when BFS is
used to sample directed networks. In the worst case scenario, if
one chooses  as a starting node a node with no outgoing links, the
procedure cannot proceed to the next step. We always choose $n=10$
seed nodes as starting nodes both to decrease the likelihood of
this type of failure and to minimize the effects of interference
between random and BFS sampling. When we sample $N$ nodes from
among the $N_0$ nodes of the real network in order to achieve a
sampling coverage, $\alpha=N/N_0$, randomly selecting $n$ nodes as
seeds affects the sampling properties of BFS, so that as $n \to
N$, BFS sampling simply becomes random sampling.

In this paper, we consider coverages of 0.25\% to 100\%. Mostly
the sampled coverage matches the target coverage as shown in
Fig.~\ref{fig:method}. However, because BFS sampling gathers new
nodes by successively exploring nodes' outgoing neighbours, its
coverage cannot exceed the combined size of the GSCC and OUT, which may relate with the ``reachability" in directed networks~\cite{Lind2007}. We
analyze all properties of the sampled network as a function of the
sampled coverage $\alpha$. For every coverage, each sampling
method was executed one hundred times on each network.

\subsection{Sampling Measurements}

We measure the following directed network properties: {\em average
degree} $\langle k \rangle$, {\em variances} of incoming and
outgoing degrees ($\sigma^2_{\rm in}$, $\sigma^2_{\rm out}$), {\em
degree auto-correlation}
 $~r_{a}$~\cite{Dorogovtsev2002,Newman2003}, {\em link reciprocity}
$R$~\cite{Garlaschelli2004}, and four kinds of {\em assortativity}
($r_{\rm ii}$, $r_{\rm io}$, $r_{\rm oi}$, and $r_{\rm
oo}$)~\cite{JFoster2010}. These will be defined in corresponding
subsections. In addition, SCC analyses are performed, and we study
how the SCCs and bow-tie structure change in response to sampling.
For each sampling coverage, we record the ratios between the sizes
of the GSCC, OUT, IN, TEND, and DISC, as well as how many nodes
comprise these components' surfaces -- i.e., their points of
contact~\cite{Donato2008,Levene2004}. We further measure how, for
BFS, the growing front ratio depends on $\alpha$. All the basic
measurements for the complete data sets are summarized in
Tables~\ref{summary1} and~\ref{summary2} as a baseline to compare
with sampling results.

\section{Sampling Results}
\label{Sampling_Results}
\subsection{Average Degree and Degree Variances}

Each node $i$ in a directed network has a number, $ k_{\rm in}^i$,
of incoming links (pointing to the node) and a number, $k_{\rm
out}^i$, of outgoing links (pointing away from the node). The
average total degree of a directed network is $\langle k\rangle =
\langle k_{\rm in}\rangle  + \langle k_{\rm out}\rangle $, where $
\langle k_{\rm in}\rangle = N^{-1} \sum_{i \in \mathbb{V}}
k^i_{\rm in} $, and similarly for  $\langle k_{\rm out}\rangle$.
Here $N$ is the number of sampled nodes in the network and
$\mathbb{V}$ is the set of sampled nodes.  Of course, $\langle
k_{\rm in}\rangle = \langle k_{\rm out}\rangle = {\langle
k\rangle}/{2}$. The variances for in- and out-degrees are
$\sigma_{\rm in}^2 = N^{-1} \sum_{i \in \mathbb{V}} (k^i_{\rm
in})^2 - \langle k_{\rm in} \rangle^2 $, and $\sigma_{\rm out}^2$
is similarly defined. Each network has a broad in-degree
distribution and a narrower out-degree distribution. Therefore all
networks exhibit a higher variance of in-degree than that of
out-degree as indicated in Table~\ref{summary2}.

For undirected networks, in the case of uniform random sampling, the sampled degree distribution
$p'(k)$ can be written as,
\begin{equation}
p'(k) = \sum_{k_0=k}^{\infty} p(k_0) {k_0 \choose k} \alpha^k
(1-\alpha)^{k_0-k}, \label{eq:pk}
\end{equation}
where $p(k)$  is the degree distribution of the original network
and $\alpha$ is the sampled
coverage~\cite{Cohen2000,Stumpf2005,Stumpf2005a,SHLee2006}.
Equation~(\ref{eq:pk}) also describes the incoming and outgoing
degree distribution of randomly sampled directed networks, where
$k$ and $k_0$ are replaced with $k_{\rm in}$ and $k_{0,\rm in}$
(or $k_{\rm out}$ and $k_{0,\rm out}$ respectively). The average
degree of the sampled network, $\langle k \rangle'$, is
\begin{eqnarray}
\langle k \rangle' &=& \sum_{k=1}^{\infty} k p'(k)
=\alpha\sum_{k_0=1}^{\infty} k_0 p(k_0) = \alpha \langle k
\rangle,
\end{eqnarray}
where $\langle k \rangle$ is the average degree of the original
network.

The variance of the degree under uniform random sampling is
obtained from
\begin{eqnarray}
\langle k^2 \rangle' &=& \sum_{k=1}^{\infty} k^2 p'(k) = \alpha^2
\langle k^2 \rangle + \alpha(1-\alpha)\langle k \rangle,
\label{eq:k2}
\end{eqnarray}
giving
\begin{eqnarray} \sigma'^2 &=& \langle k^2
\rangle' - \left(\langle k \rangle'\right)^2 = \alpha^2 \langle
k^2 \rangle + \alpha(1-\alpha)\langle k \rangle - \alpha^2 \langle
k \rangle^2 \nonumber\\
&=&\alpha^2 \sigma^2 + \alpha(1-\alpha)\langle k \rangle
\label{eq:sigma}
\end{eqnarray}
where $\sigma^2$ represents the variance in degree of the original
network. The same formulas, Eqs.~(\ref{eq:k2}) and
(\ref{eq:sigma}), also hold for the variances of the in- and
out-degree, respectively. Thus $\sigma_{\rm in}'^2$ and
$\sigma_{\rm out}'^2$ are both quadratic functions of $\alpha$,
although with different coefficients ($\sigma_{\rm in}^2 \neq
\sigma_{\rm out}^2$). When the coverage $\alpha$ is small,
$\sigma'^2$ increases linearly with $\alpha$; for large $\alpha$
it increases quadratically.

\begin{figure}[b]
\begin{flushleft}
\includegraphics[width=0.42\textwidth]{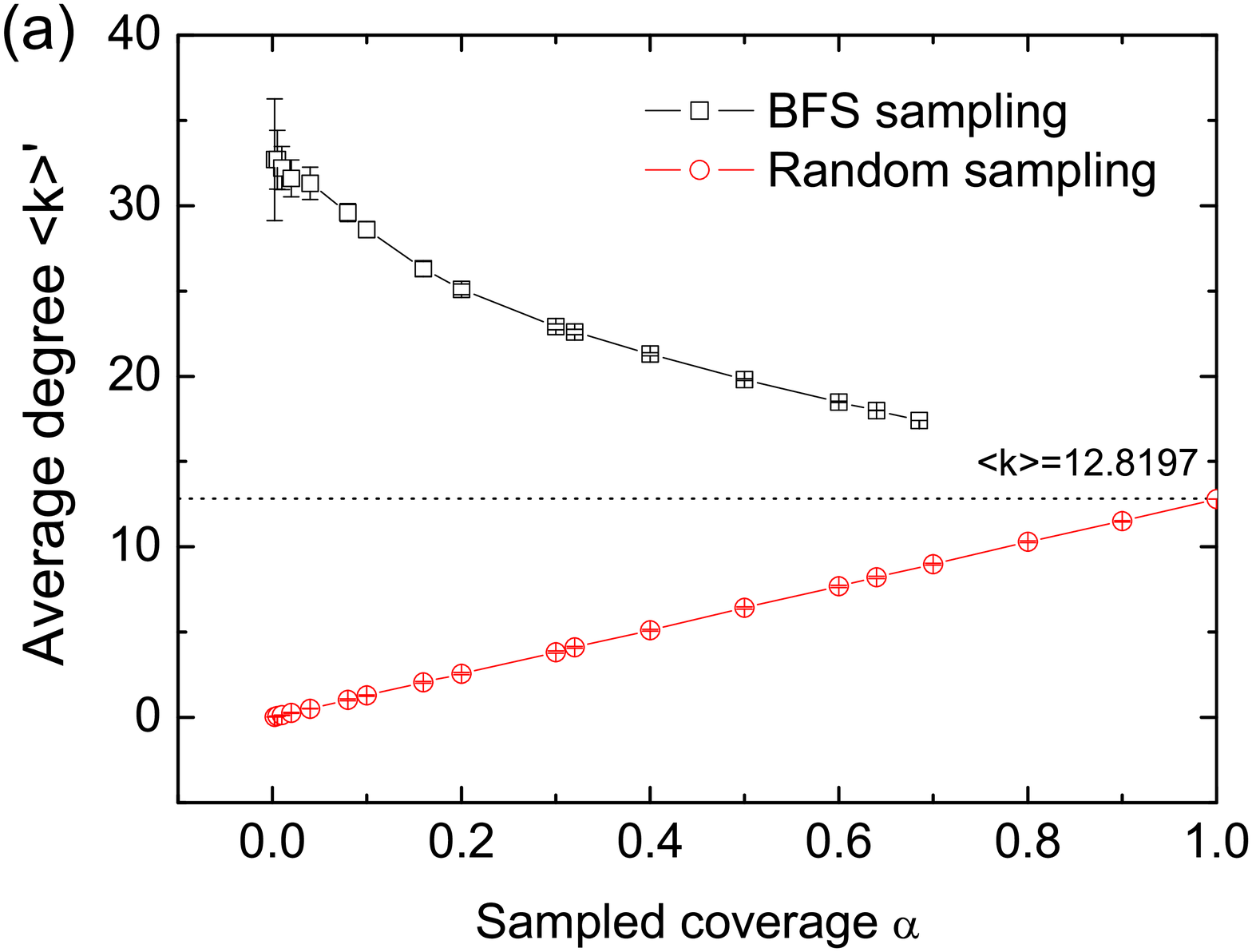}
\includegraphics[width=0.46\textwidth]{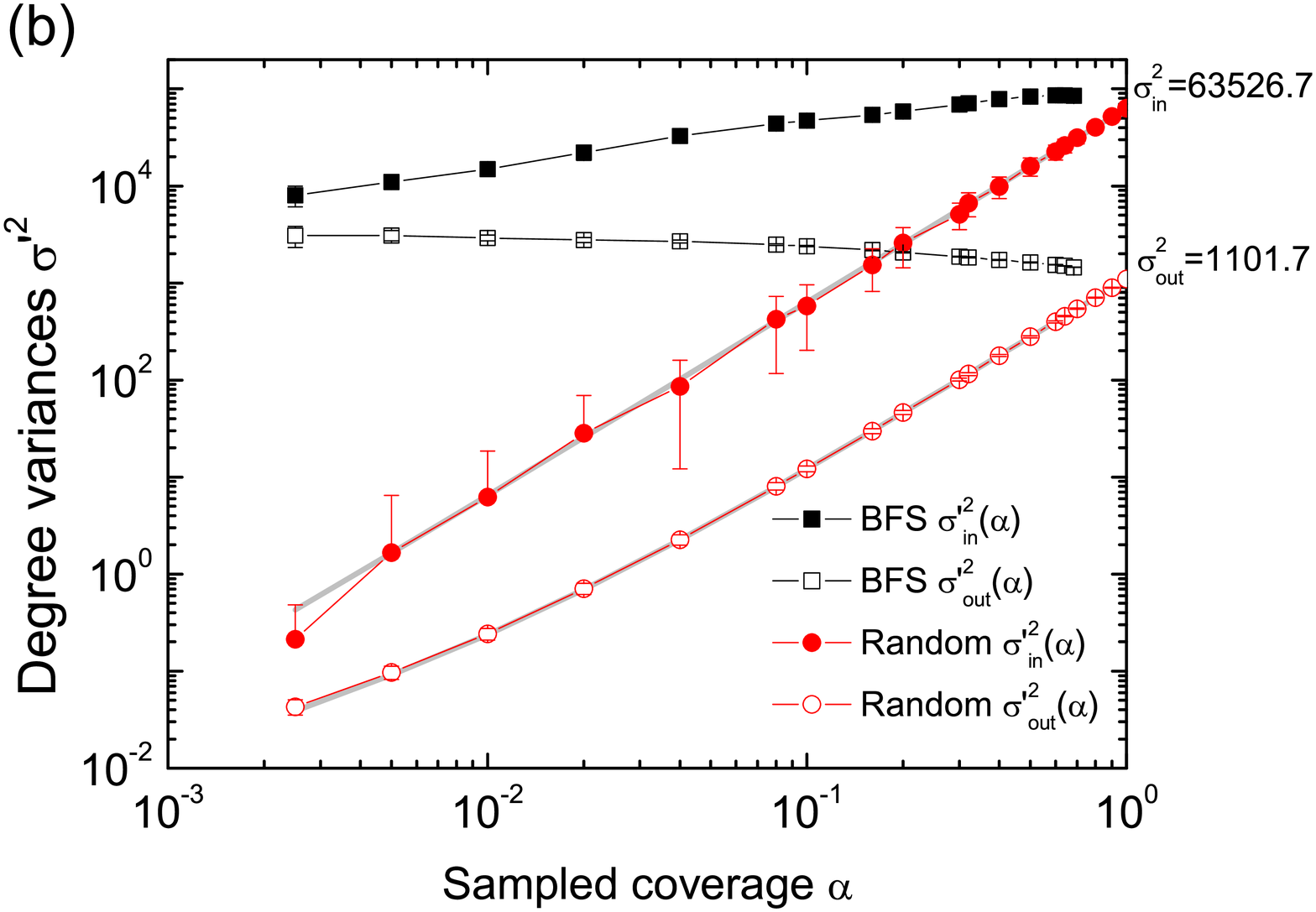}
\caption{ (Color online) Behaviors of  the average degree (a) and
 the variance of the in-degree and
out-degree (b) as a function of sampling coverage, $\alpha$,  for
each sampling method in the Wiki2007 data. (a) In the case of BFS
sampling, the average degree approaches its asymptotic value-- the
average degree of the combined GSCC and OUT-- from above, since
BFS is biased to the high degree nodes. The average degree for
random sampling is just linearly proportional to the sampling
coverage. (b) The variances for in- and out-degrees for random
sampling increase quadratically as sampled coverage increases, but
those of BFS approach their real values from opposite directions.
The gray lines behind the random sampling data are calculated from
Eq.~(\ref{eq:sigma}). Hereafter all the error bar means the
standard deviation of the measured variables over sampling
realizations.} \label{fig:degree}
\end{flushleft}
\end{figure}

This quadratic relation is shown for Wiki2007 in
Fig.~\ref{fig:degree}(b). The gray lines behind the random
sampling data indicate the results calculated from
Eq.~(\ref{eq:sigma}). Since the variance of the incoming degree is
much larger than the average degree, $\sigma_{\rm in}'^2$ seems to
be purely quadratic in this plot, but the variance of the outgoing
degree, $\sigma_{\rm out}'^2$, shows the transition from a linear
to a quadratic function as $\alpha$ increases. As can be seen in
Fig.~\ref{fig:degree}(b), random sampling severely underestimates
variances of in- and out-degree (by as much as two orders of
magnitude even at $\sim 10\%$ coverage). This underestimation
results from the quadratic dependence of Eq.~(\ref{eq:sigma}) on
$\alpha$.

\begin{figure}[t!]
\includegraphics[width=0.35\textwidth]{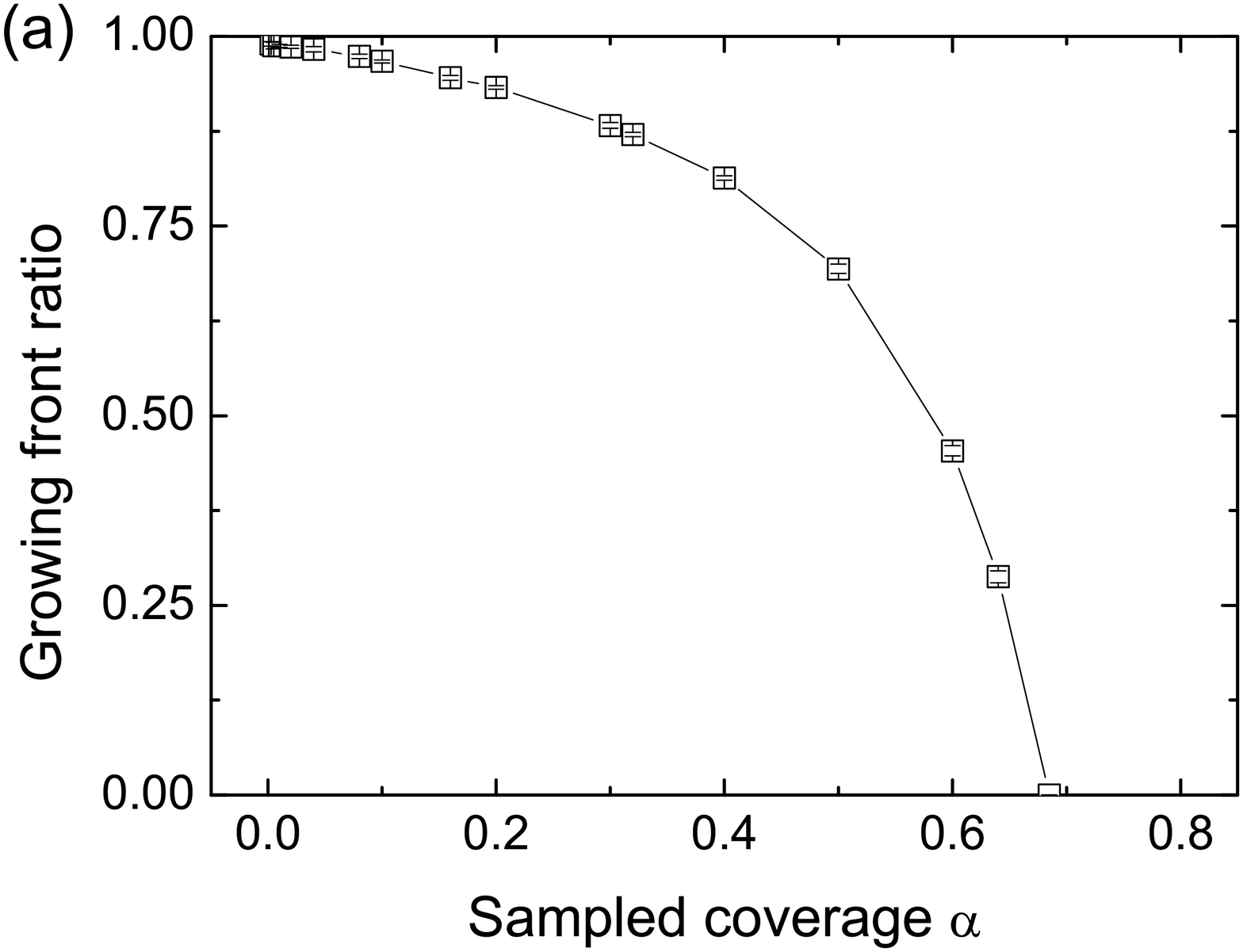}
\includegraphics[width=0.35\textwidth]{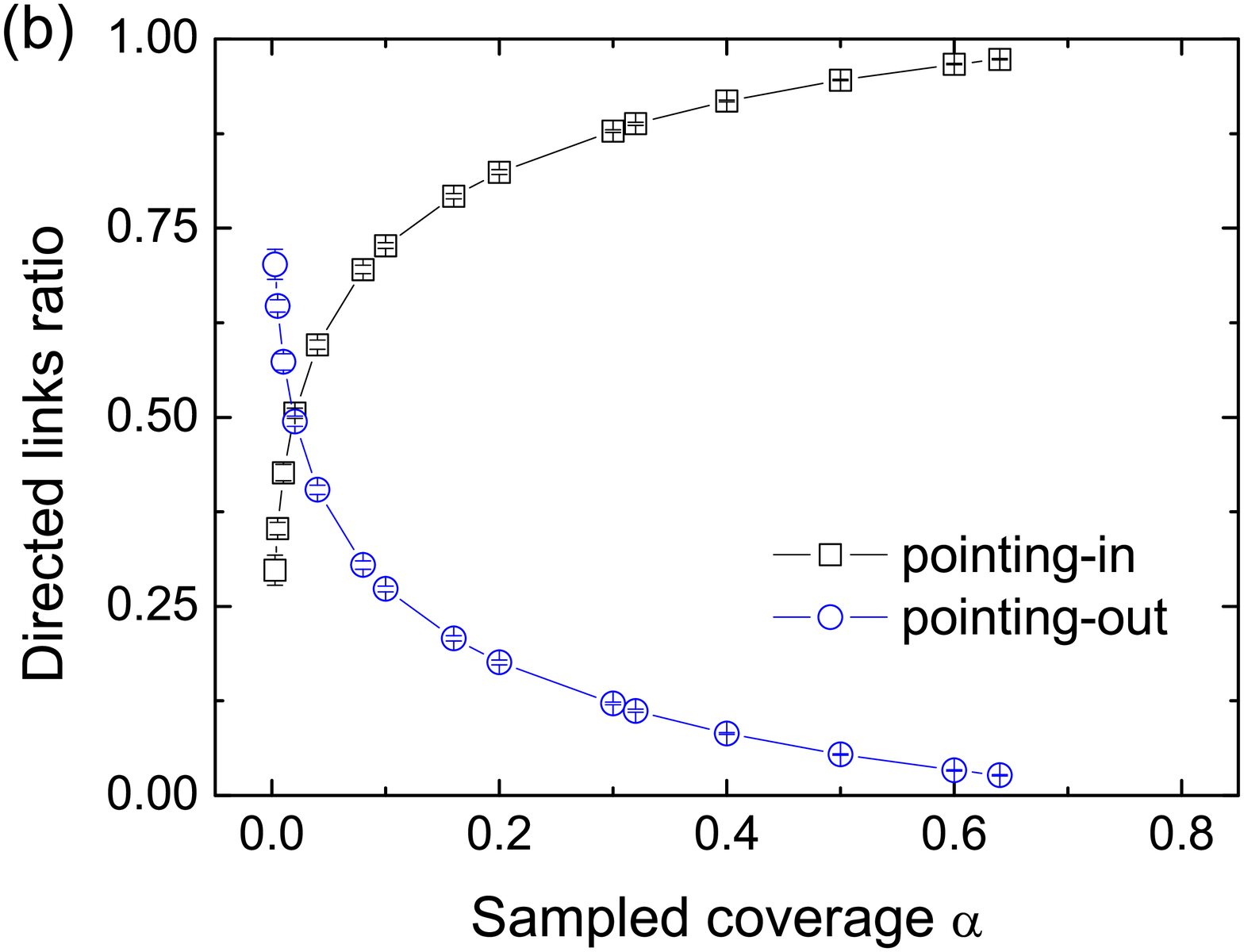}
\caption{ (Color online) For BFS sampling of Wiki2007 data, (a)
growing front ratio as a function of coverage and (b) directed
links ratio (the fraction of edges in the  growing front that
point outside (circles) or back into (squares) the already sampled
network). Note that even for small coverage a substantial fraction
of links directing from the growing front point back to the
already sampled network. } \label{fig:gfront}
\end{figure}

As shown in Fig.~\ref{fig:degree}, BFS sampling does not obey
these simple mathematical relationships. Since BFS follows
outgoing links and reaches hub nodes at early
times~\cite{SHLee2006,Becchetti2006}, one could have an argument
that the average degree of BFS-sampled networks overestimates the
average degree. However, this can only be true when the networks
contain loops. In the case of a tree, since the network resulting
from BFS sampling is still a tree, the average degree is
$2-\frac{2}{N}$ very close to $\langle k \rangle = 2-\frac{2}{N_0}
\approx 2$. The average degree of BFS-sampled networks is related
to the loop structures and clustering. Therefore we measure the
size of the growing front under BFS sampling and the number of
directed links pointing into the already sampled networks as shown
in Figs.~\ref{fig:gfront}(a) and (b). In early stages of BFS
sampling, although most nodes lie in the growing front, the
fraction of their links pointing back to the already sampled nodes
is surprisingly high.

BFS sampling also overestimates variance of out-degree, but
underestimates variance of in-degree as can be seen in
Fig.~\ref{fig:degree}(b). However, these errors are less severe
than for random sampling. Variance of in-degree is underestimated
in BFS sampling for the same reason it is underestimated in random
sampling, although the misestimations are less severe since the
correlated loop structures affects the directed link ratio of the
growing front as shown in Fig.~\ref{fig:gfront}(b). Variance in
out-degree is overestimated for a different reason: visited nodes
have the same out-degree in the sampled networks as they do in the
original networks, while the out-degree of growing front nodes is
not fully counted. As $\alpha$ increases, the effect of the
growing front nodes diminishes. Indeed the fraction of directed
(and unexplored) links pointing outside of the sampled network
shrinks quickly even though the fraction of nodes on the growing
front decreases much more slowly, as shown in
Fig.~\ref{fig:gfront}(a) and (b).

\subsection{Degree Auto-correlation}

\begin{figure}[b!]
\includegraphics[width=0.35\textwidth]{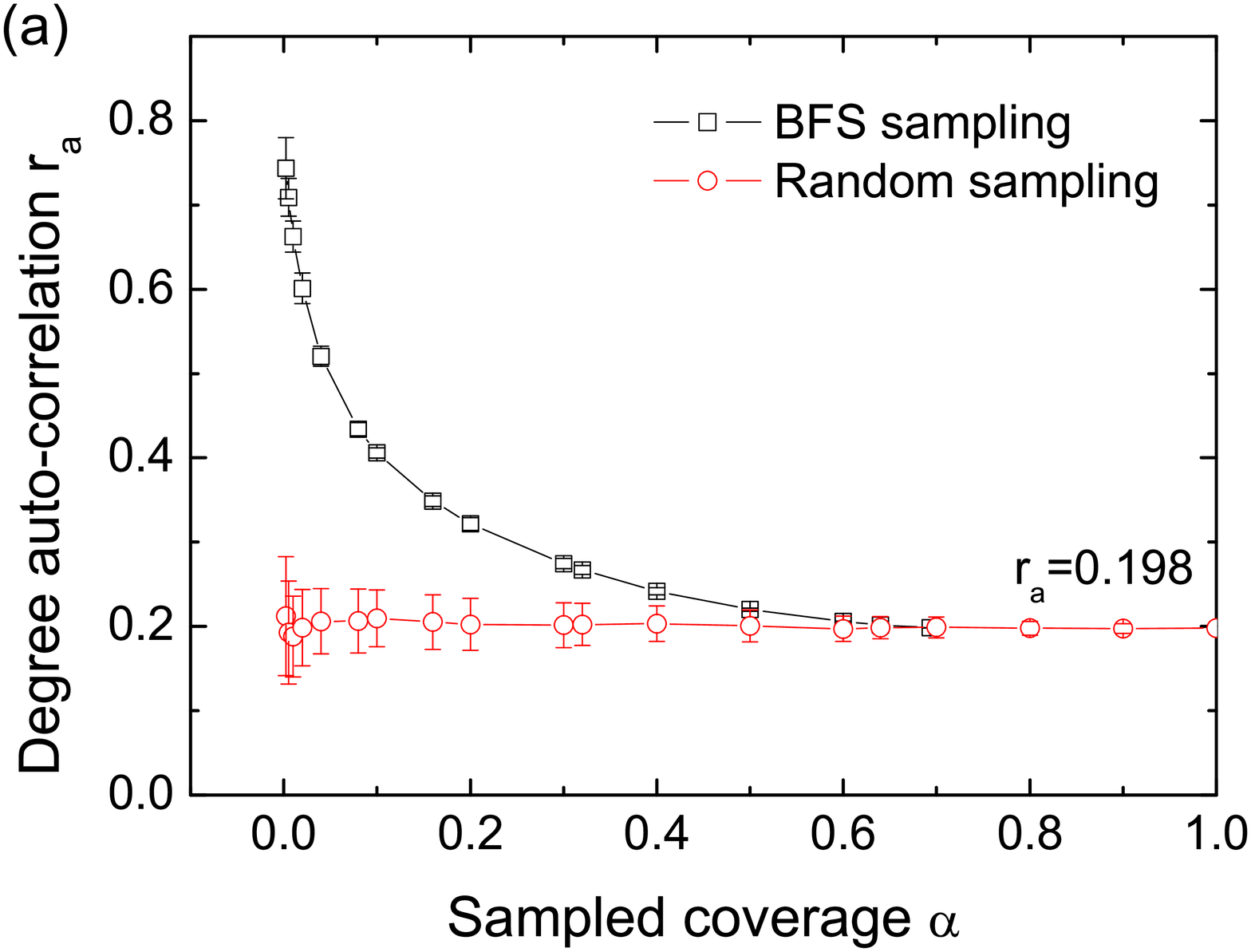}
\includegraphics[width=0.35\textwidth]{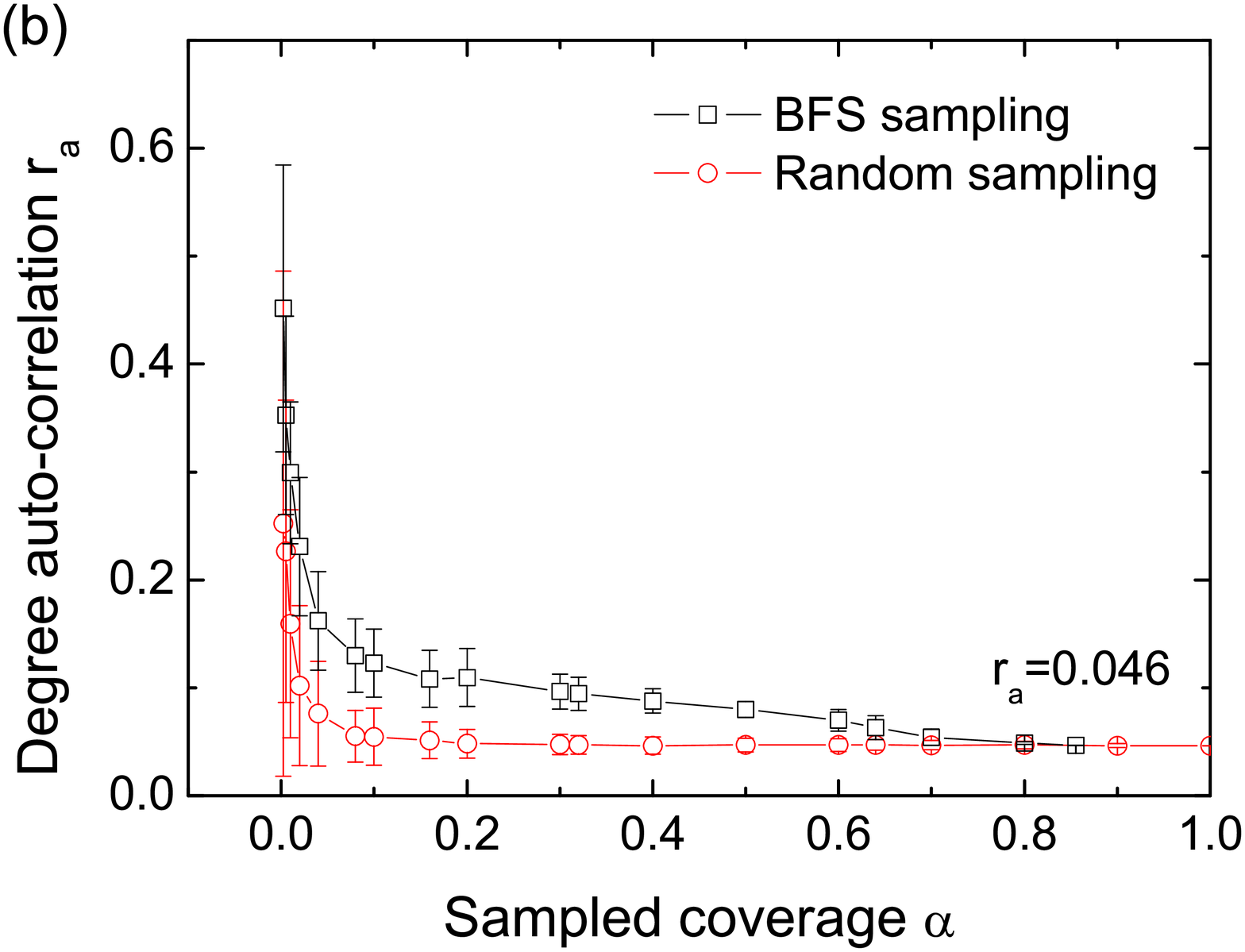}
\caption{ (Color online) Degree auto-correlation for (a) Wiki2007
data and (b) Stanford data under both sampling methods. BFS
sampling significantly overestimates this quantity.}
\label{fig:auto-correlation}
\end{figure}

\begin{table*}[t]
\begin{minipage}{0.8\textwidth}
\caption{\label{degree_reciprocity} Summary of incoming and
outgoing degree and average local reciprocity for each component.
Here $\langle \cdot \rangle$ means the average over nodes in each
component.}
\centering%
\rowcolors{3}{}{gray!35}
\renewcommand{\tabcolsep}{0cm}
\renewcommand{\arraystretch}{1.2}
\begin{tabular}{lrrrrrrrr}
\hline
\multirow{2}*{~Component~~~~} & \multicolumn{4}{c}{\footnotesize Wiki2007} & \multicolumn{4}{c}{\footnotesize Google} \\
& \scriptsize size (\%) & \scriptsize $\langle k_{\rm in} \rangle$ & \scriptsize $\langle k_{\rm out} \rangle$ & \scriptsize $\langle R_i \rangle$ & \scriptsize ~~~~~size (\%) & \scriptsize $\langle k_{\rm in} \rangle$ & \scriptsize ~~~$\langle k_{\rm out} \rangle$ & \scriptsize $\langle R_i \rangle $ \\
\hline
~GSCC & 67.2 & ~~~18.72 & ~~~17.76 & ~~0.122  & 50.8 & ~~~9.51 & 8.40 & ~~0.330 \\
~OUT & 1.4 & 15.88 & 0.04 & ~~0.0004 & 19.4 & 3.46 & 1.35 & 0.292 \\
~IN & 31.4 & 0.11 & 2.84 & 0.006  & 21.1 & 1.47 & 5.87 & 0.188 \\
~TEND & 0.1 & 1.06 & 0.44 & 0.090 & 8.7 & 1.25 & 1.76& 0.190 \\
\hline
~GWCC & 100.0 & \multicolumn{2}{c}{~~~12.82} & 0.118 & 100.0 & \multicolumn{2}{c}{~~~5.92} & 0.306 \\
\hline
\end{tabular}
\end{minipage}
\end{table*}

Degree auto-correlation quantifies the extent to which nodes of
high in-degree also have high out-degree, and is defined as $r_{a}
= {\rm Cov}(k_{\rm in}, k_{\rm out})/\sigma_{\rm in}\sigma_{\rm
out}$. The covariance is given by, ${\rm Cov}(k_{\rm in}, k_{\rm
out}) = N^{-1} \sum_{i \in \mathbb{V}} k_{\rm in}^i k_{\rm out}^i
- \langle k_{\rm in}\rangle \langle k_{\rm out} \rangle$.  All
networks, except for BerkStan and Stanford, have moderately high
degree auto-correlation ($r_{a}
>0.1$).

In the case of random sampling, the degree auto-correlation,
$r_a$, is unbiased if $\alpha$ is large enough to ensure an
adequate density of links, since the in- and out-degrees for each
node are sampled randomly. Figures~\ref{fig:auto-correlation}(a)
and (b) show this effect, although, for small $\alpha$, some nodes
are isolated and therefore have no in- or out-degree, trivially
causing an increase in degree auto-correlation
(Fig.~\ref{fig:auto-correlation}(b)). As can also be seen in
Figs.~\ref{fig:auto-correlation}, BFS enhances -- by up to 400\%
-- degree auto-correlation at low sampling coverage.

\subsection{Reciprocity}

The link reciprocity R is defined as the fraction of links in a
network that participate in a two-way relationship, {\em i.e.}, $R
\equiv L_{\leftrightarrow} / L$, where $L_{\leftrightarrow}$ means
the number of edges belonging to bidirectional connections and $L$
is the total number of links in the
network~\cite{Garlaschelli2004}. For each node $i$, we can also
similarly define a local recicprocity $R_i$, which is the fraction
of node $i$'s edges belonging to bidirectional connections.

\begin{figure}[b!]
\includegraphics[width=0.40\textwidth]{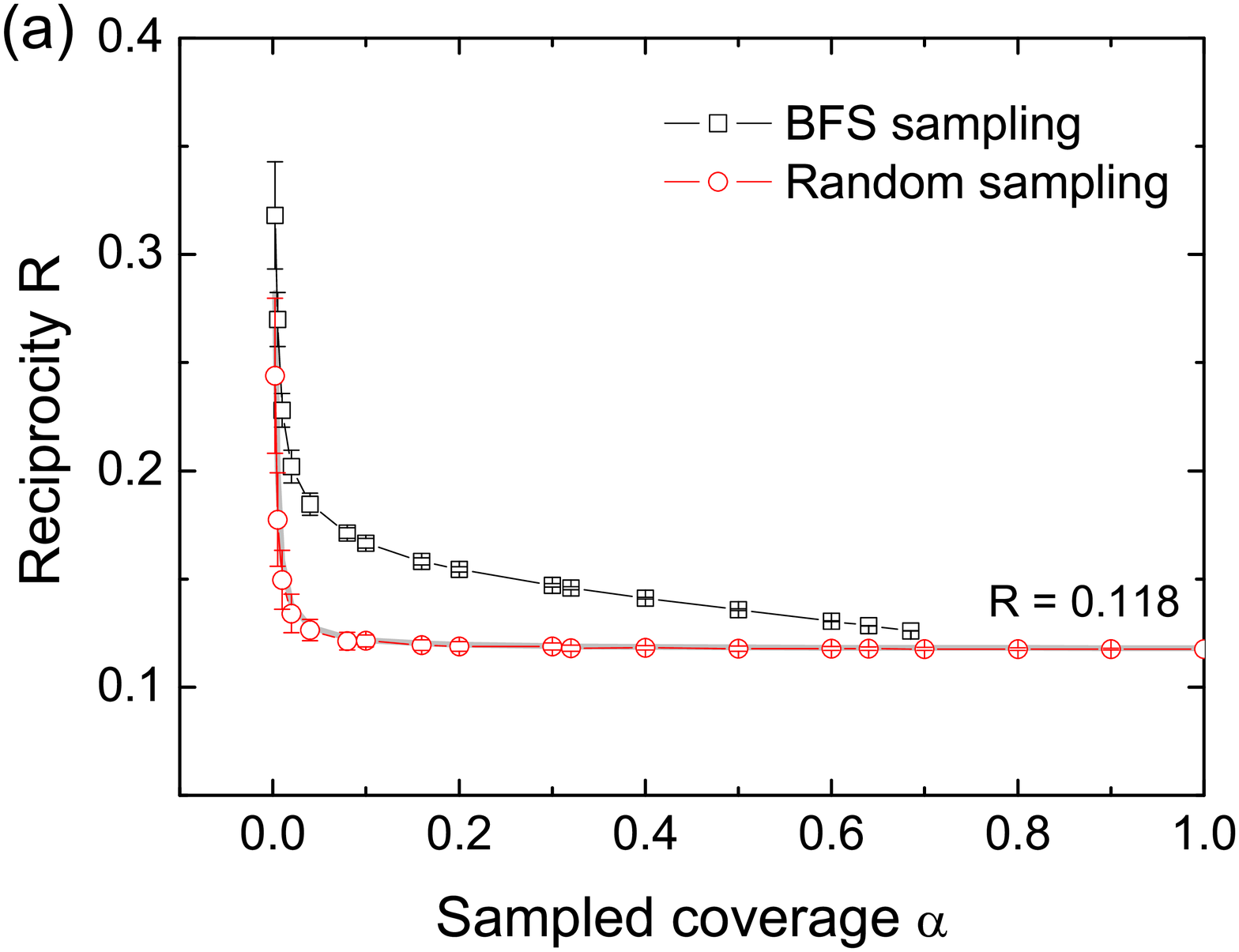}
\includegraphics[width=0.40\textwidth]{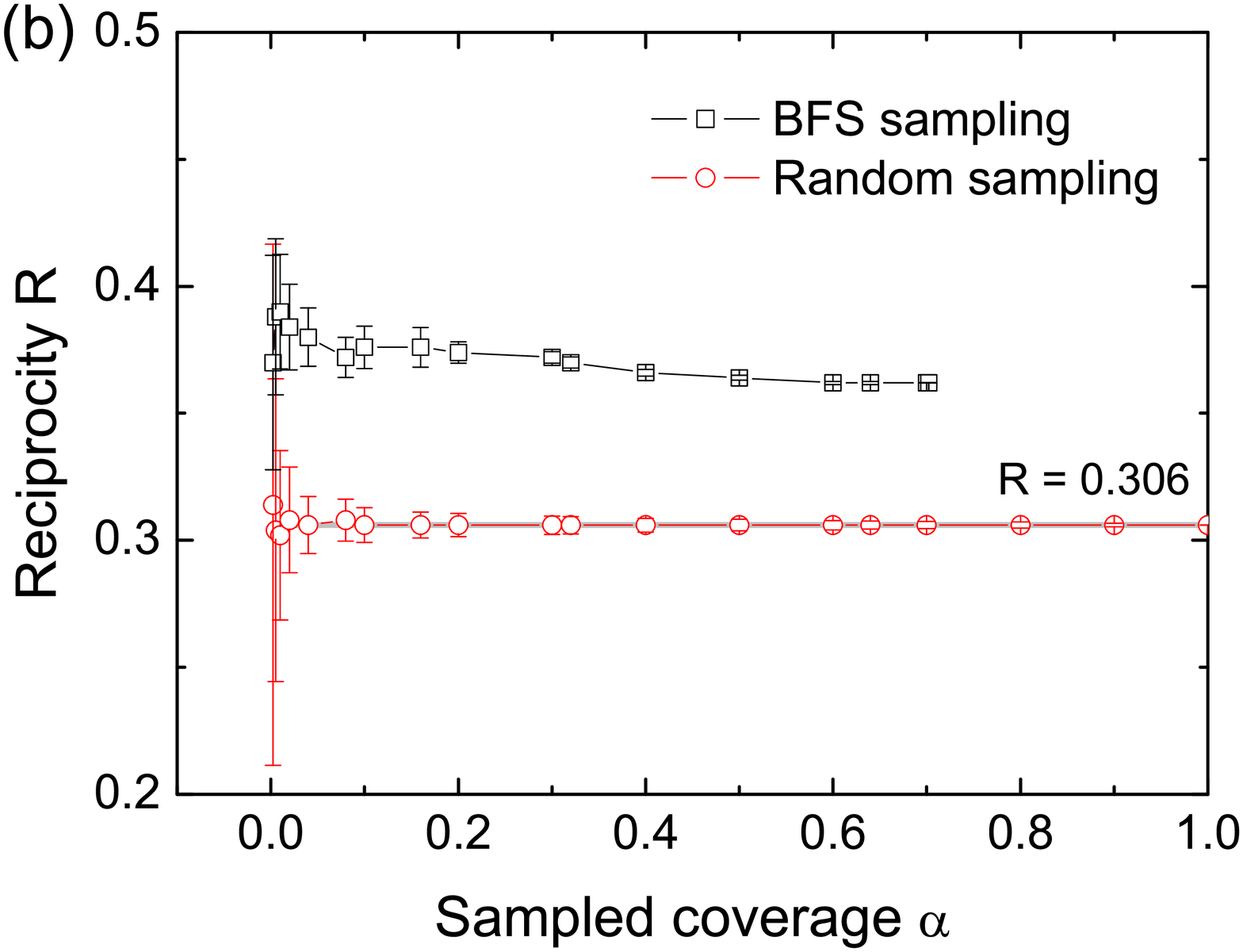}
\caption{ (Color online) Reciprocity changes as sampled coverage
increases for Wiki2007 (a) and Google (b) data. Except for an
initial state that results from the presence of self-links, random
sampling shows constant reciprocity while BFS sampling approaches
the true value slowly from above. The gray lines behind the random
sampling data are the theoretical prediction for random sampling.}
\label{fig:R}
\end{figure}

For random sampling, in the absence of self-links ({\em i.e.}
links that start and end on the same node), reciprocity is
constant, independent of sampling coverage, since any pair of
nodes is chosen with the same probability as any other pair of
nodes. If, however, self-links are present, the reciprocity under
random sampling is higher than that of the true network since the
self-links (which are reciprocal by definition) appear with
probability $\alpha
> \alpha^2$. The reciprocity with respect to $\alpha$ is $R
(\alpha) \approx R (1 + \alpha^{-1} c/L_{\leftrightarrow} )$,
where $c/L_{\leftrightarrow}$ is the fraction of self-links among
bidirectional links. Thus, one can see that in the presence of
self-links, the reciprocity is no longer constant, but quickly
approaches its asymptotic value as ${\alpha}$ increases.  The data
in Fig.~\ref{fig:R}(a) illustrate this effect for Wiki2007 which
exhibits a small fraction ($< 0.4\%$) of self-links among all
bidirectional links. The gray lines in the figures are the
expectation lines from the above equation and agree perfectly with
the data.

For BFS sampling, however, reciprocity is significantly
overestimated, and only slowly approaches its true value. At least
part of the bias for reciprocity under BFS comes from the fact
that we only include links in the growing front if they point back
to the previously sampled graph. This introduces a bias to
increase reciprocity. This type of overestimation is actually
present at any sampling coverage for an additional reason: BFS
sampling (in most cases) only gathers information about the GSCC
and OUT components, but not about the IN and other components, and
since reciprocal links always tie two nodes into one component,
there are naturally more reciprocal links in the GSCC than there
are in other components as summarized in
Table~\ref{degree_reciprocity}; thus there is overrepresentation
of bidirectional links, relative to the total number of links, and
reciprocity is artificially high as shown most clearly in
Fig.~\ref{fig:R}(b) for the Google data.

\begin{figure*}[t!]
\centering
\includegraphics[width=0.40\textwidth]{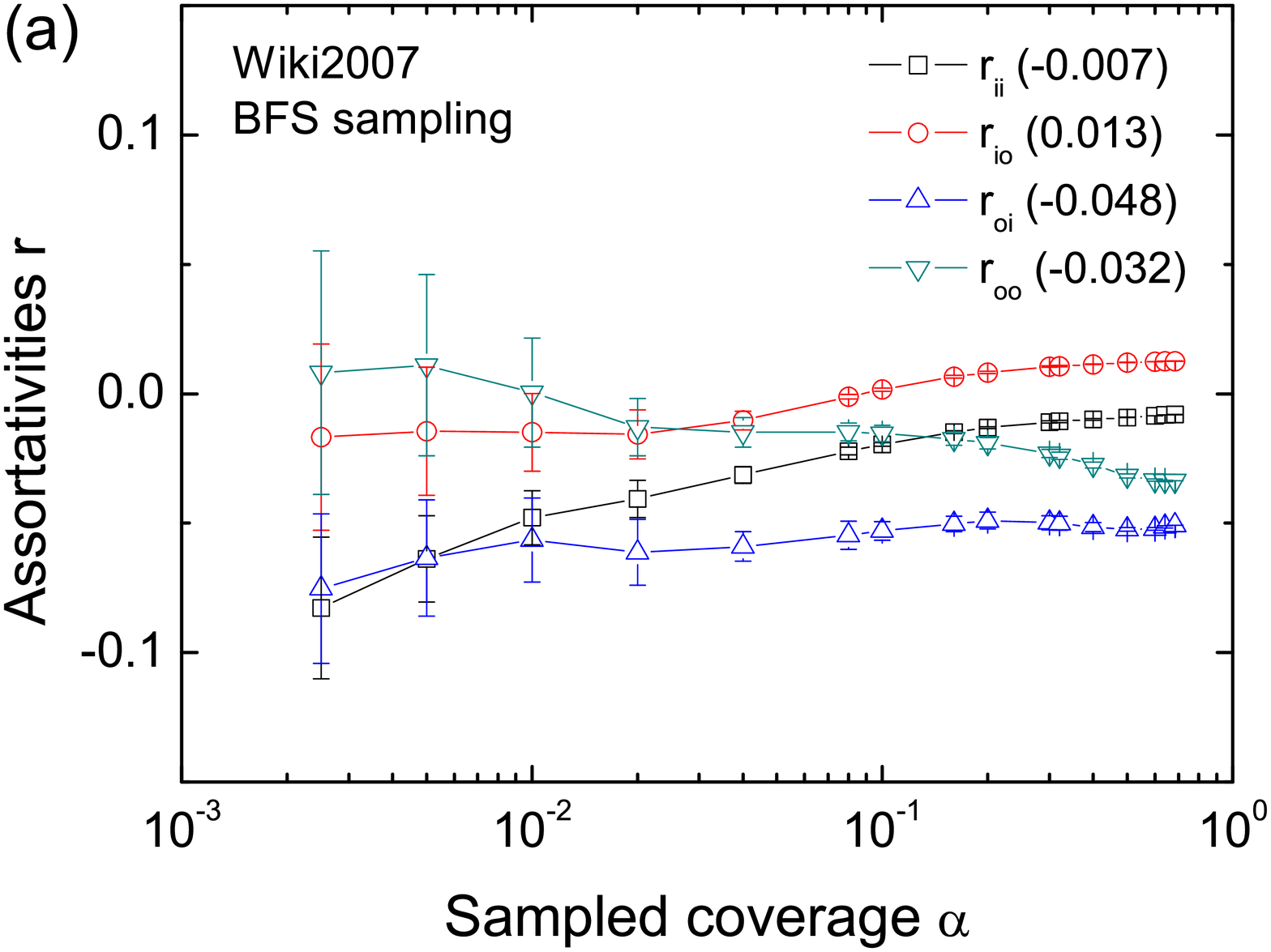}
\includegraphics[width=0.40\textwidth]{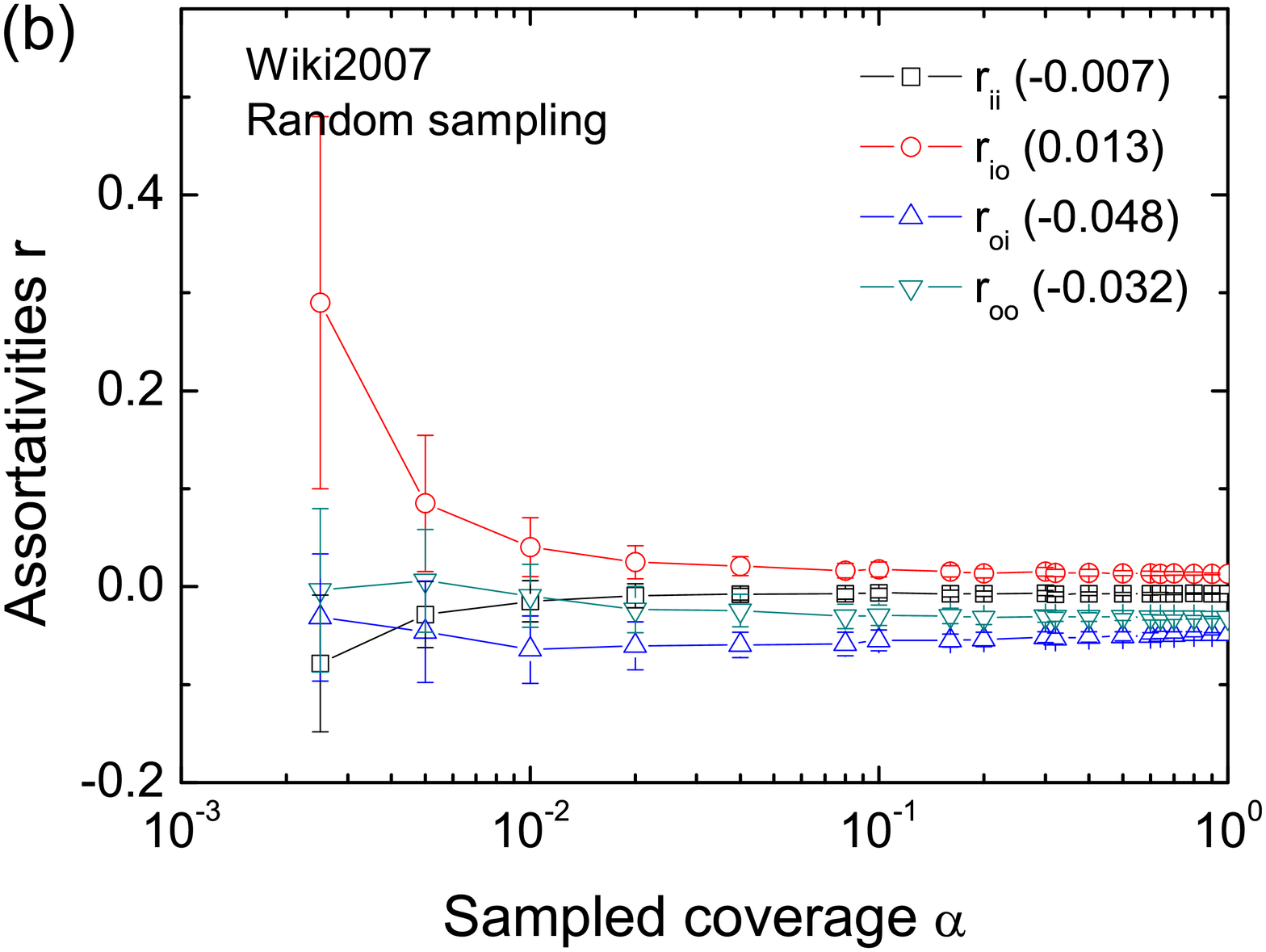}
\includegraphics[width=0.40\textwidth]{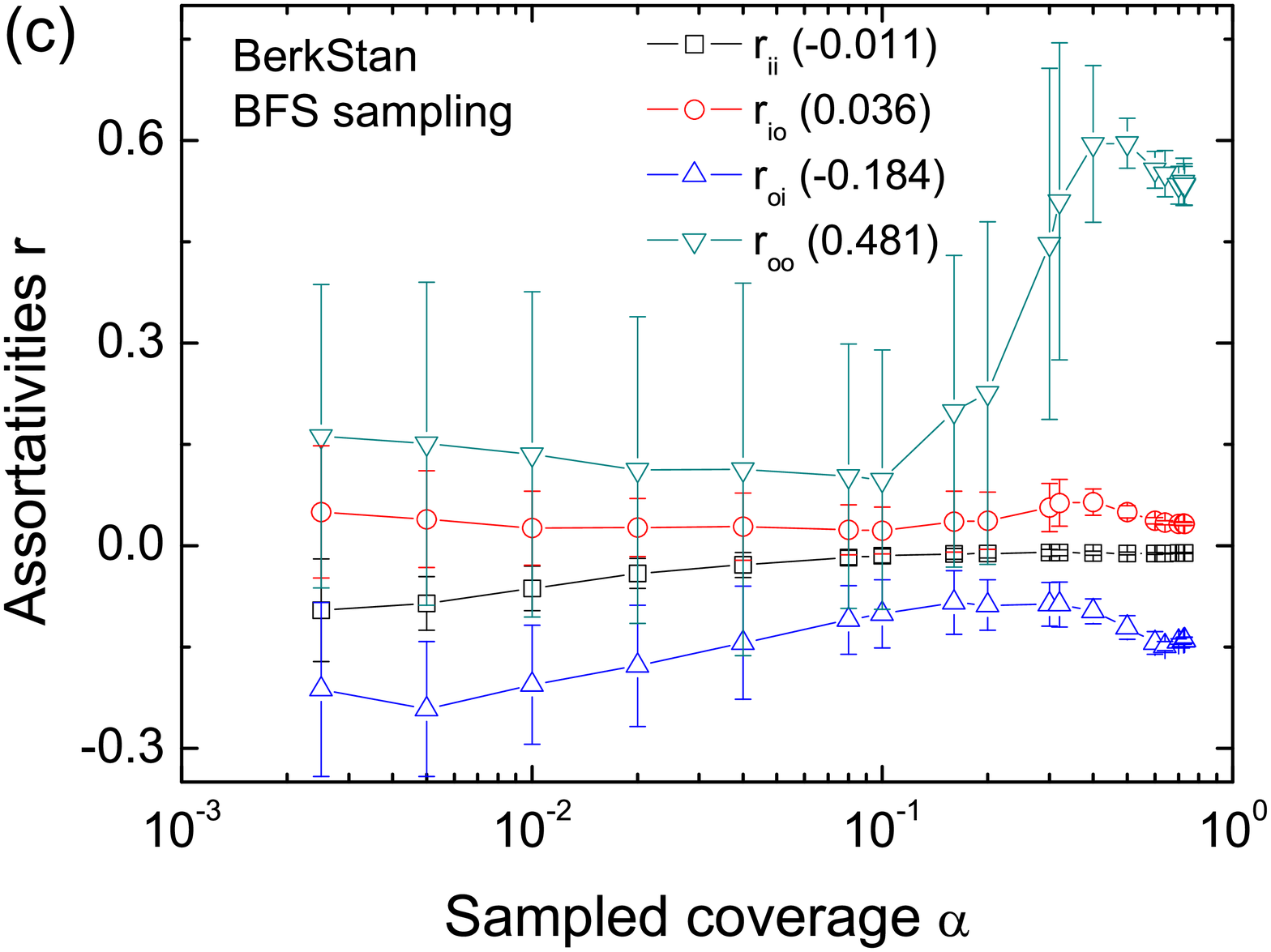}
\includegraphics[width=0.40\textwidth]{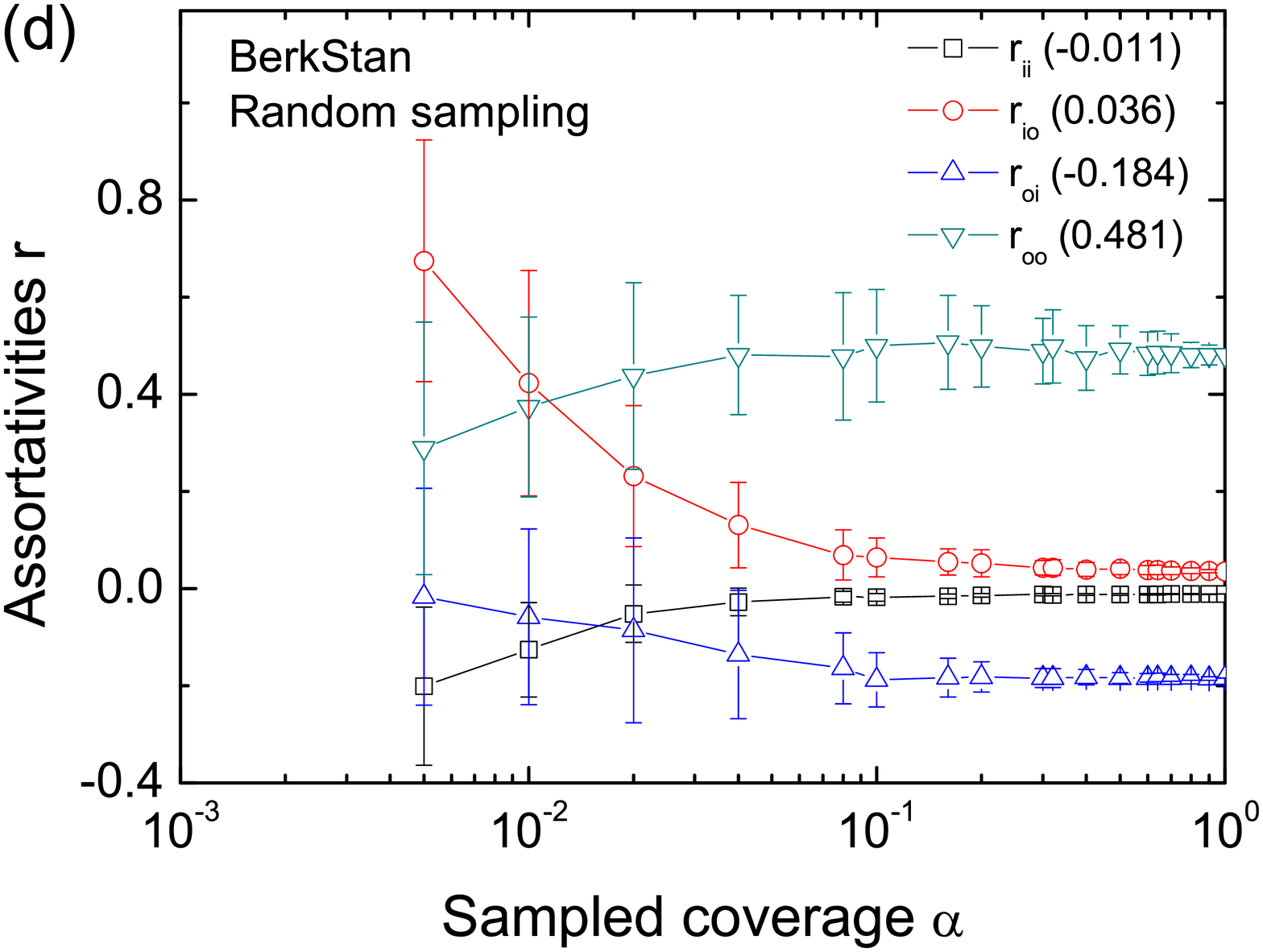}
\caption{ (Color online) Four kinds of assortativity for
BFS-sampled networks and random-sampled ones on (a), (b) Wiki2007
data and (c), (d) BerkStan data.} \label{fig:Assortativity}
\end{figure*}

\subsection{Assortativities}

A set of assortativity measures~\cite{JFoster2010} for directed
networks are defined using the Pearson correlation as follows:
\begin{equation}
r_{xy} = \frac{L^{-1} \sum\limits_{l \in \mathbb{E}} \left[
(j_{x}^l - \overline{j}_{x})(k_{y}^l - \overline{k}_{y}) \right]
}{s^{(j)}_{x} s^{(k)}_{y}}, \label{eq:assortativity}
\end{equation}
where $x, y \in \{{\rm in},{\rm out}\}$ indexes each incoming and
outgoing degree type and $j_{x}^l$ ($k_{y}^l$) is the $x$-degree
($y$-degree) of the {\em tail} ({\em head}) node for a link $l$,
and $\mathbb{E}$ is the set of sampled links (all links, if we
consider the complete network). $\overline{j}_{x} = L^{-1} \sum_{l
\in \mathbb{E}} j_{x}^l$ ($\overline{k}_{y} = L^{-1} \sum_{l \in
\mathbb{E}} k_{y}^l$) is the weighted average degree~\cite{note1}. The following
relations hold in general: $\overline{j}_{\rm out} \neq
\overline{j}_{\rm in} = \overline{k}_{\rm out} \neq
\overline{k}_{\rm in}$. $s^{(j)}_{x} = \sqrt{L^{-1} \sum_{l \in
\mathbb{E}} (j_{x}^l - \overline{j}_{x})^2}$ is the standard
deviation of the $x$-degree of the tail nodes ($\neq
\sigma'_{x}$). $s^{(k)}_{y}$ is similarly defined. It is worth
noting that $s^{(j)}_{\rm in} \neq s^{(j)}_{\rm out} \neq
s^{(k)}_{\rm in} \neq s^{(k)}_{\rm out}$.

In most cases, we find that the directed assortativities of the
networks we study are not markedly different from zero, and it is
therefore difficult to define a general tendency for the effects
of BFS sampling on the statistics of assortativity.  We do,
however, point out that both $r_{\rm oo}$ and $r_{\rm oi}$ of the
BerkStan network are quite large (but have opposite sign),
suggesting that, unlike the rest of the networks, nodes of high
out-degree tend to link to other nodes of high out-degree, but
nodes of high out-degree tend to link with nodes of low in-degree.

While the small assortativities of the networks make it dangerous
to draw broad conclusions regarding the effects of sampling, it is
clear that in the case of random sampling there is a clear
tendency in behavior at low values of the sampling coverage, which
seems to be related to the small reciprocity of the networks we
study~\cite{ZamoraLopez2008}. Assortativity between the incoming
degree and outgoing degree $r_{\rm io}$ tends to be overestimated
for small sampling coverage; on the other hand, the
incoming-incoming degree assortativity is underestimated (implying
greater disassortativity than is present in the complete networks)
as shown in Figs.~\ref{fig:Assortativity}(b) and (d). These trends
seem to stem from a trivial situation:  when the sampling coverage
is small, many tail (head) nodes will have no incoming (outgoing)
degree, even though they are connected to each other. Consider two
nodes, A and B, connected by a directed link from A to B. In this
case, A has no incoming degree and B has no outgoing degree. Thus
the correlation between in- and out-degrees would be positive,
whereas the correlation between incoming degrees would be
negative. This would not be the case if a large fraction of nodes
had reciprocal links. Not surprisingly, these tendencies disappear
very quickly as the sampling coverage increases.

Assortativity can be either overestimated or underestimated by
BFS, depending on network structure and coverage. The real value
of in-degree/in-degree assortativity is approached from below in
Wikipedia data (see Fig.~\ref{fig:Assortativity}(a)). When
randomly picking the seeds for BFS, there is a high chance to
select small $k^i_{\rm in}$ nodes since incoming degree follows a
scale-free distribution. Nonetheless, BFS sampling soon reaches
the large $k^i_{\rm in}$ nodes. This results in a highly negative
$r_{\rm ii}$ initially. As the sampling coverage increases,
$r_{\rm ii}$ approaches its real value from below. However, we do
not observe systematic behaviors for other assortativities under
BFS.

\subsection{Number of SCCs}

\begin{figure}[b!]
\includegraphics[width=0.36\textwidth]{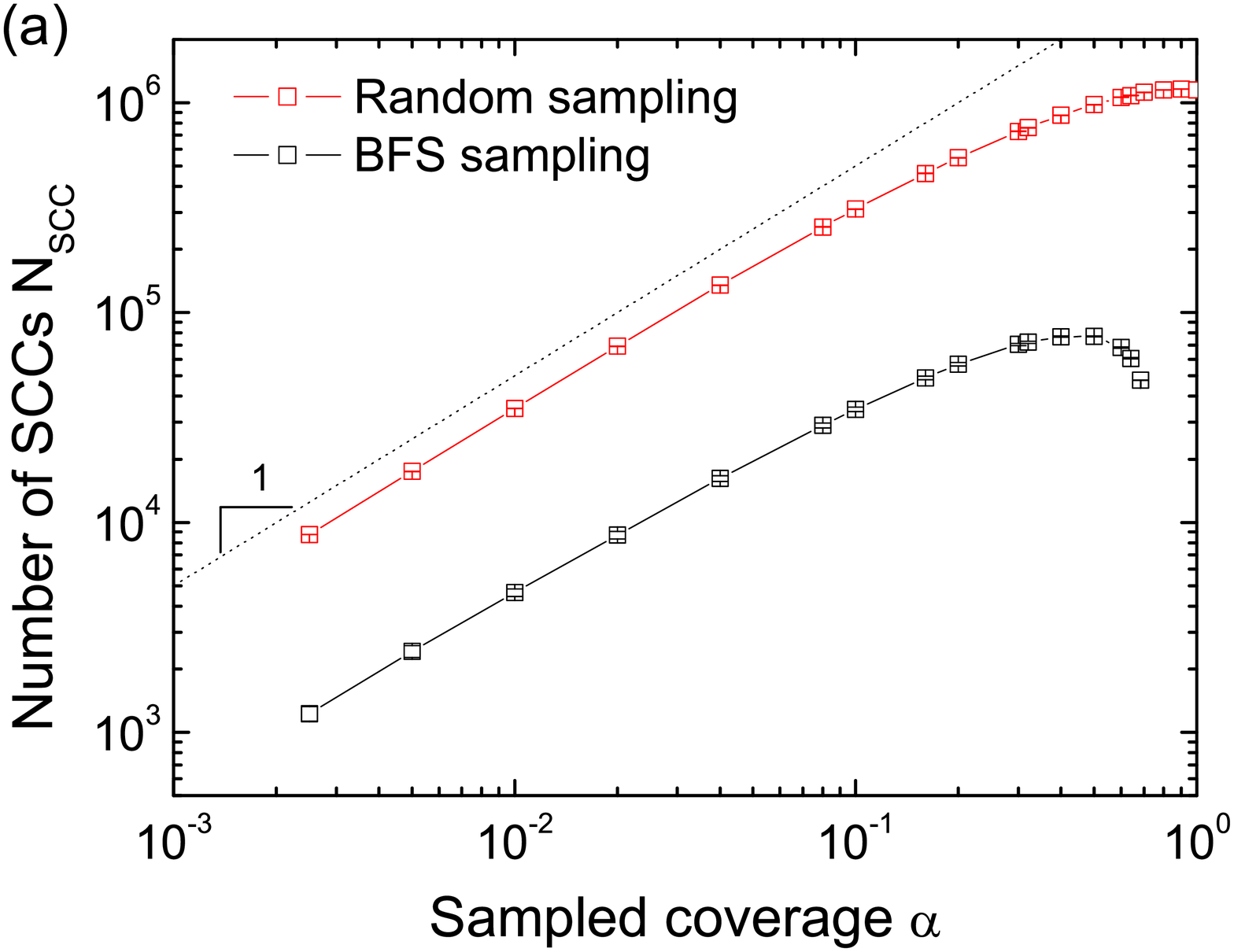}
\includegraphics[width=0.36\textwidth]{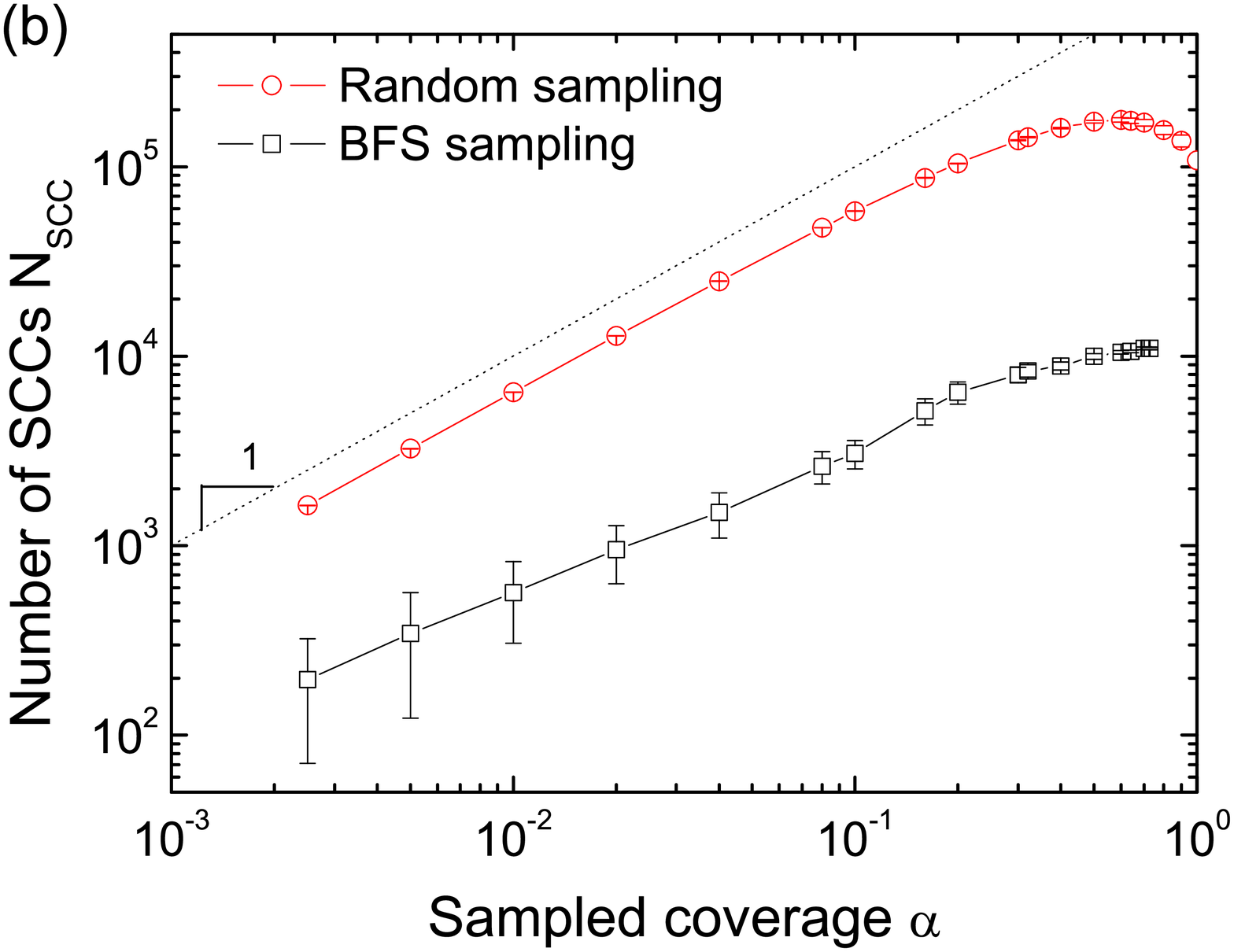}
\caption{ (Color online) Tendency for the number of SCCs in
Wiki2007 (a) and BerkStan (b) data to increase with both sampling
methods. After a threshold sampling coverage has been reached, the
number of SCCs will decrease, since the newly-sampled nodes will
bridge preexisting SCCs. The black dotted lines in (a) and (b) are
the reference for the slope 1.} \label{fig:SCCs}
\end{figure}

As the sampling coverage increases, the number of SCCs increases
initially. Since single nodes and nodes with only incoming links
are considered SCCs by definition, the number of SCCs is
proportional to the sampling coverage $\alpha$, both for random
and BFS sampling. However, after a certain sampling coverage has
been reached, newly-sampled nodes are more likely to connect to
already-existing SCCs.  For most networks, this means that
existing SCCs will merge together, whence the total number of SCCs
will finally decrease. This is illustrated in Fig.~\ref{fig:SCCs}.
For both sampling methods, the number of SCCs increases linearly
with $\alpha$ initially and then decreases to the  value in the
original networks for large $\alpha$. However, the number of SCCs
observed in BFS sampling is almost one order of magnitude less.

\subsection{Surface Nodes}

\begin{figure}[b!]
\includegraphics[width=0.36\textwidth]{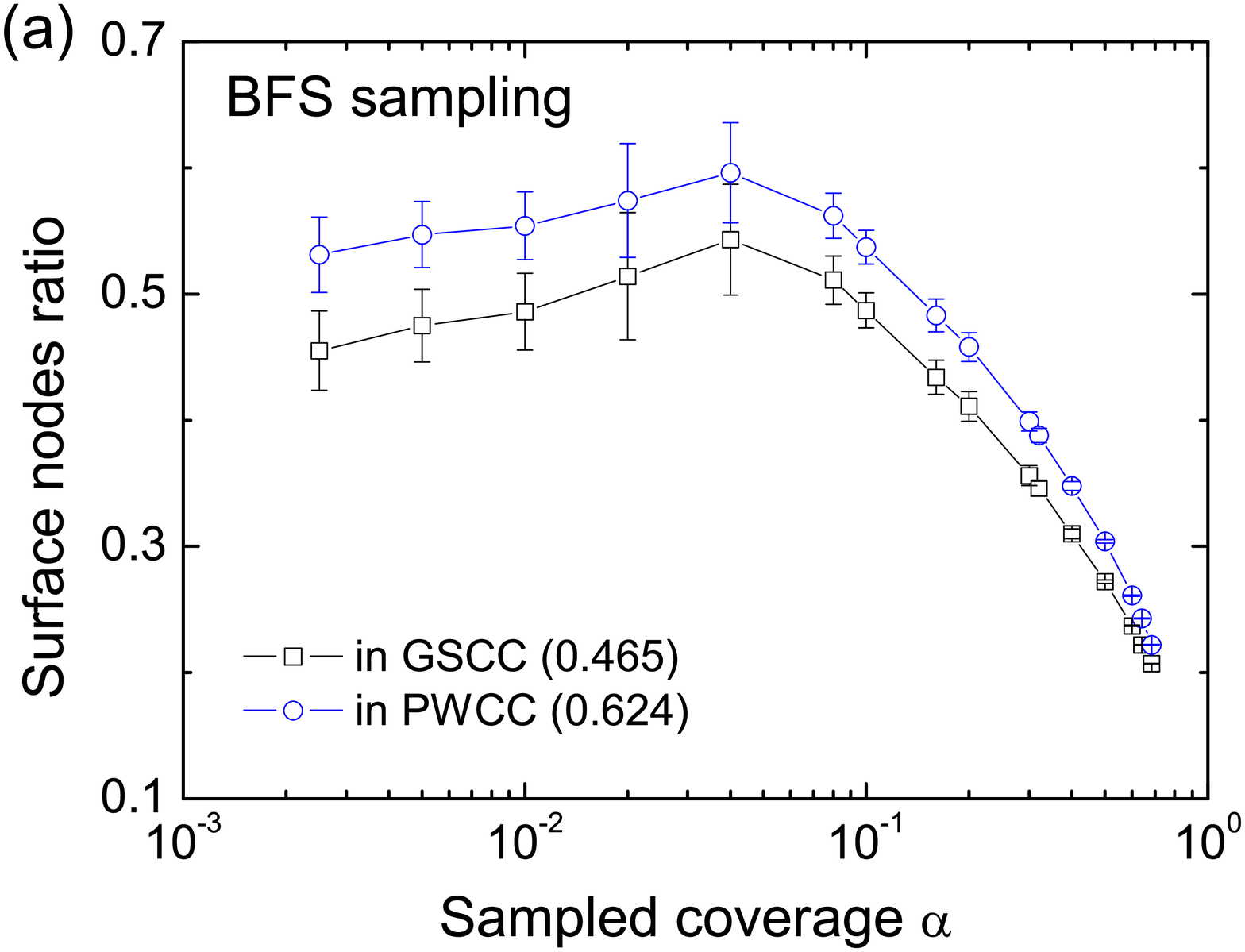}
\includegraphics[width=0.36\textwidth]{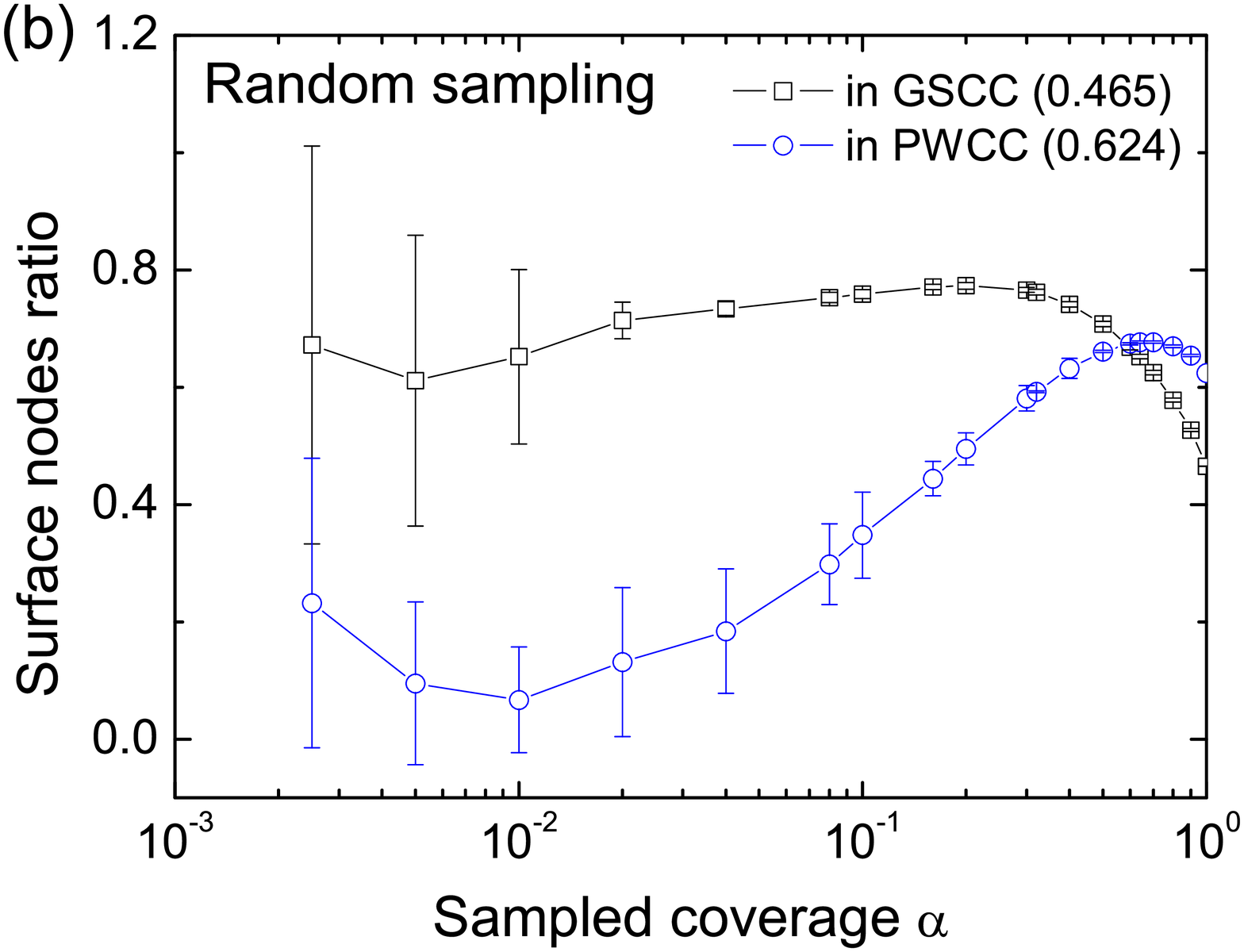}
\caption{ (Color online) Ratio of the surface nodes in the GSCC
and the PWCC of (a) BFS-sampled and (b) randomly sampled networks
of Wiki2007 data. In the case of the BFS sampling, the estimated
values do not approach to the true values since BFS sampling only
covers the all nodes in GSCC and OUT components.}
\label{fig:Surface}
\end{figure}

Since surface nodes are in contact with other components, there is
a possibility that they will be absorbed into component cores or
move into other components if we add nodes or links from the
network. The ratio of nodes on the surface of a component to the
total number of nodes in the component (`surface node ratio')
seems to depend strongly on the structure of the SCCs of directed
networks.  Of the networks we study, the Wikipedia graphs have the
largest GSCCs, with upwards of 67\% of all nodes, and at least
43\% of theses nodes are surface nodes.  The Stanford and BerkStan
networks' GSCCs are smaller (59\% and 51\%, respectively) and
contain very few surface nodes (7.4\% and 9.6\%, respectively).  A
closer look at the IN and OUT components of the Stanford network
reveals numerous chains and multinode (directed) cycles that offer
only a single surface node for attachment to the GSCC.
Figure~\ref{fig:Surface} shows the changes to surface node ratios
as sampling coverage increases in the Wiki2007 data.

When the sampling coverage is small, the surface node ratio in the
GSCC does not change markedly under random sampling. After
increasing the sampling coverage, however, the ratio decreases as
the core becomes more densely connected with the addition of newly
sampled nodes. However, the surface node ratio in the PWCC
increases as shown in Fig.~\ref{fig:Surface}(b) as the DISC and
TEND shrink quickly, becoming absorbed into the GSCC, and then
transforming into surface nodes.

BFS sampling, on the other hand, shows a different trend. The
surface node ratio in the GSCC is lower than that in the PWCC.
This seems to be deeply related with the fact that BFS sampling
starts from seeds than expands their territory layer by layer.
When the sampling coverage is small, the surface node ratio
increases as the sampling coverage increases. After the sampling
procedure has reached a certain point, the surface node ratio will
also begin to decrease as shown in Fig.~\ref{fig:Surface}(a).

\begin{figure*}[th!]
\centering
\includegraphics[width=0.40\textwidth]{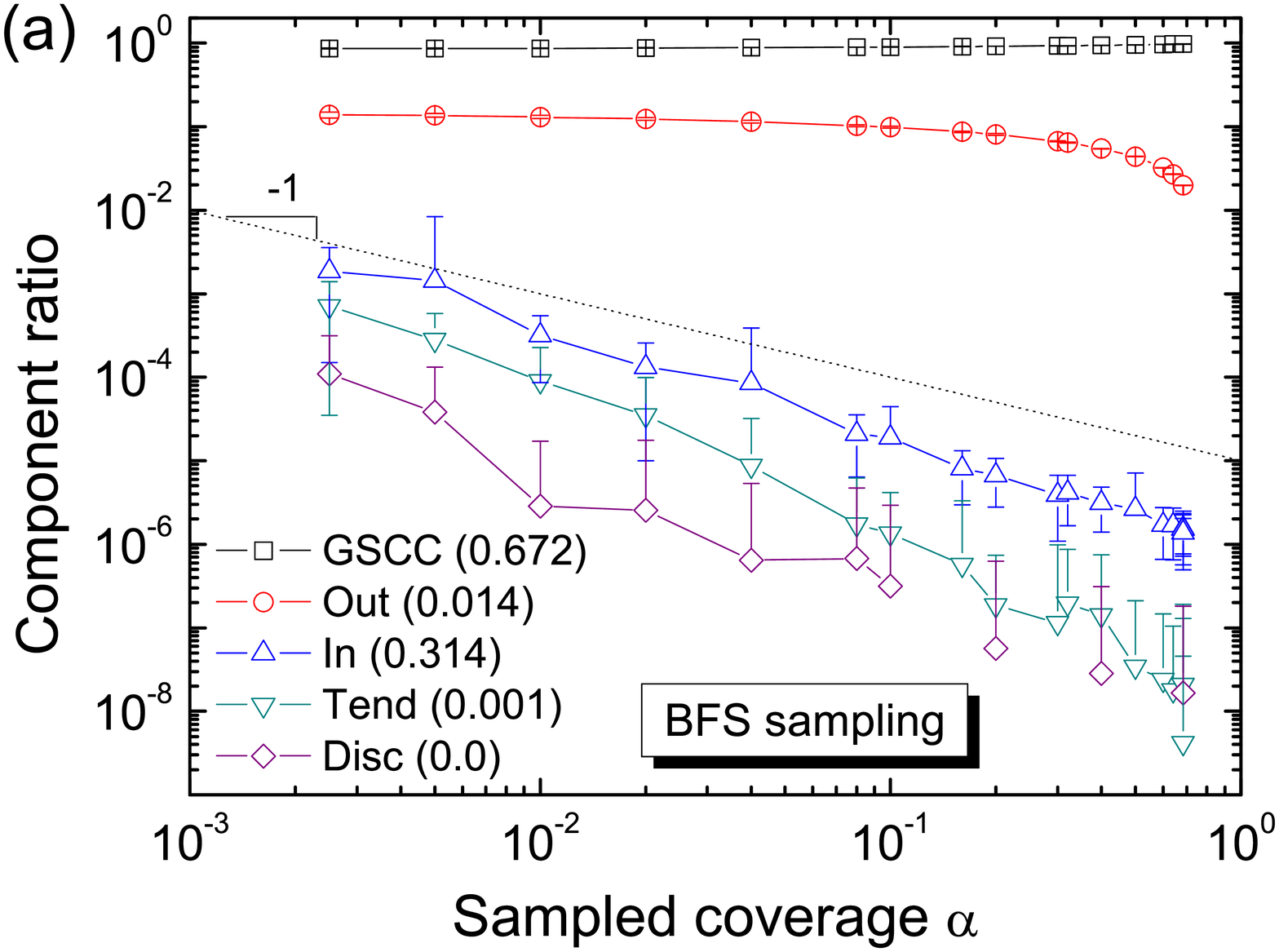}
\includegraphics[width=0.40\textwidth]{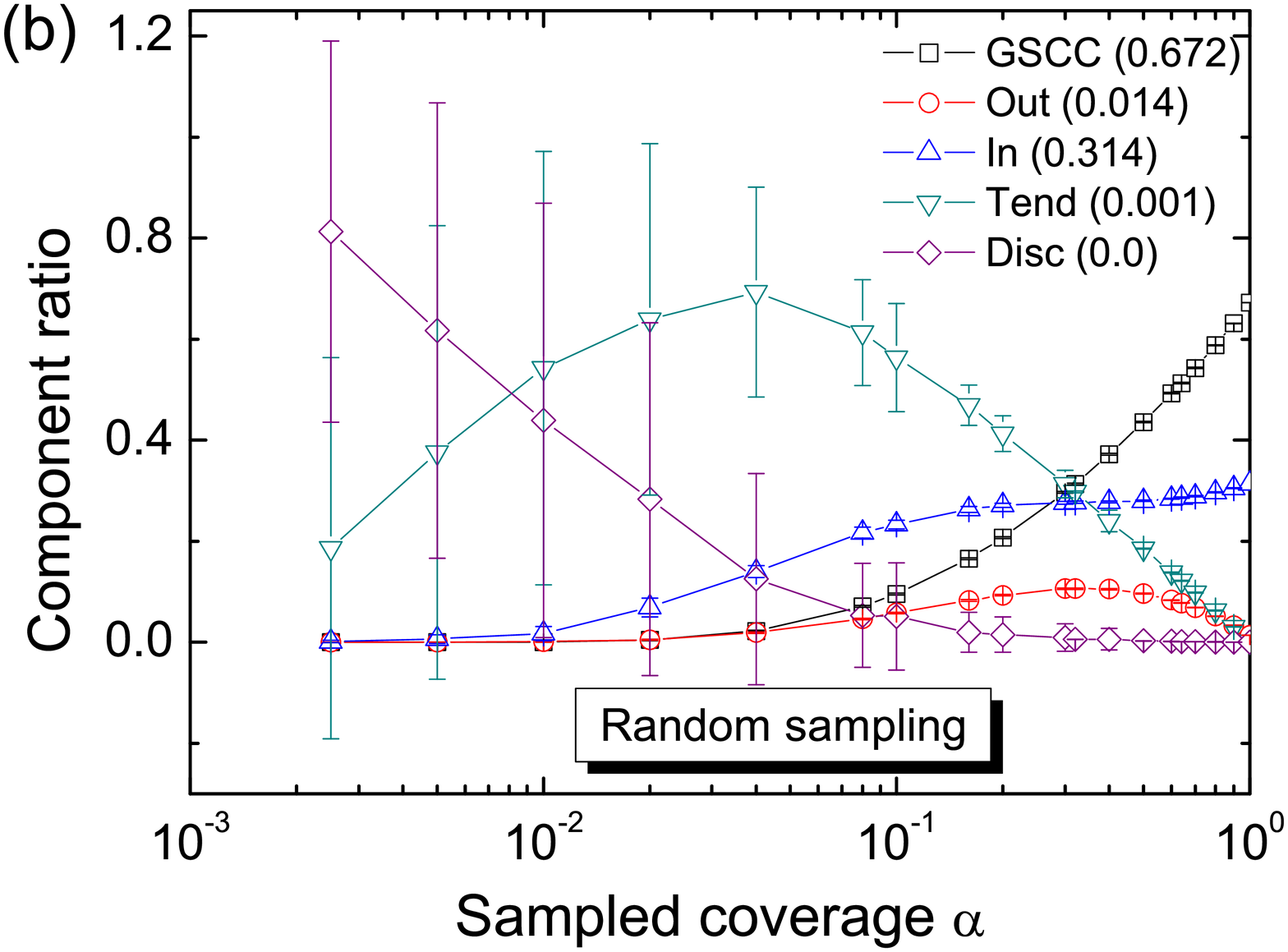}
\includegraphics[width=0.40\textwidth]{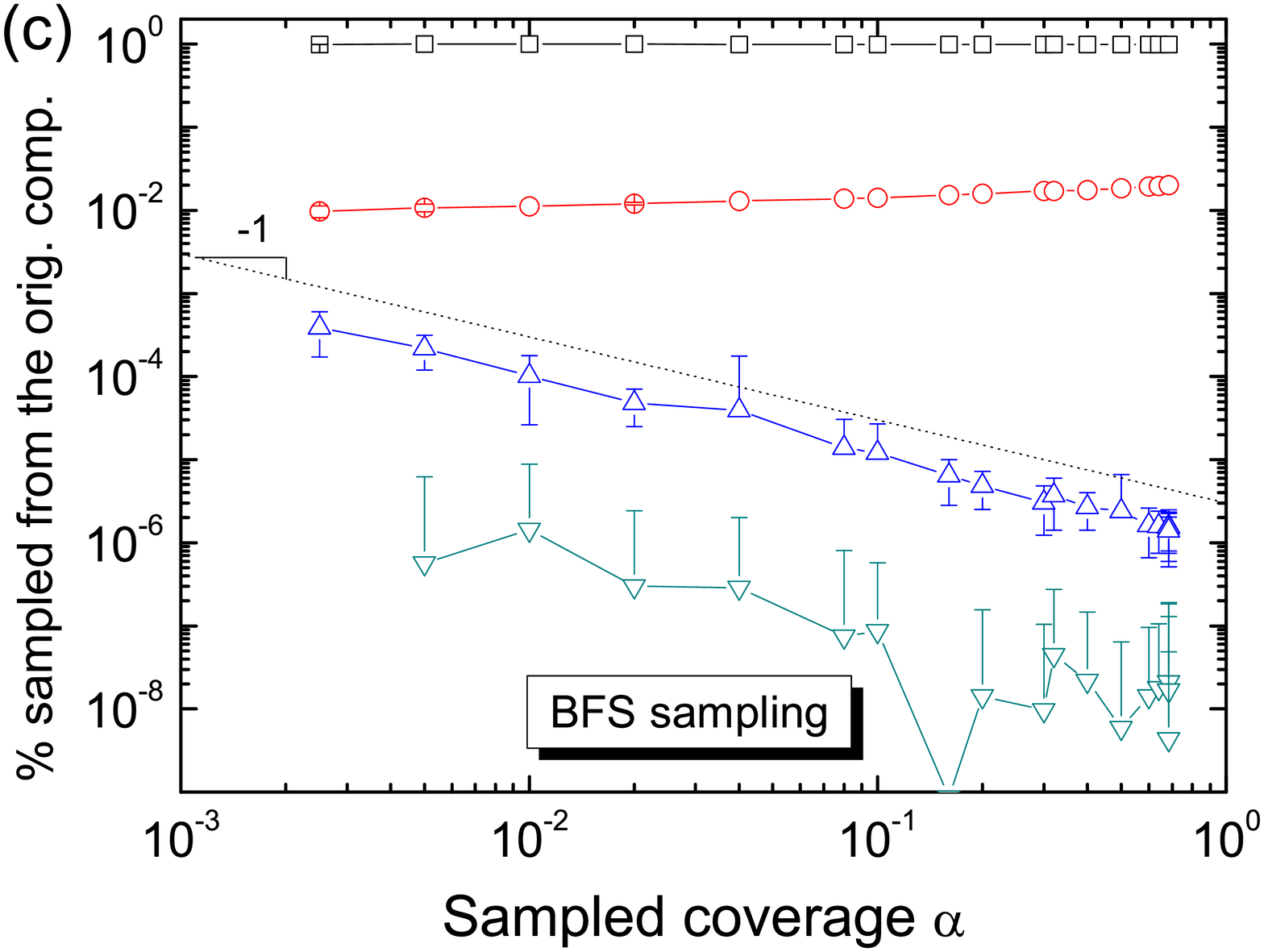}
\includegraphics[width=0.40\textwidth]{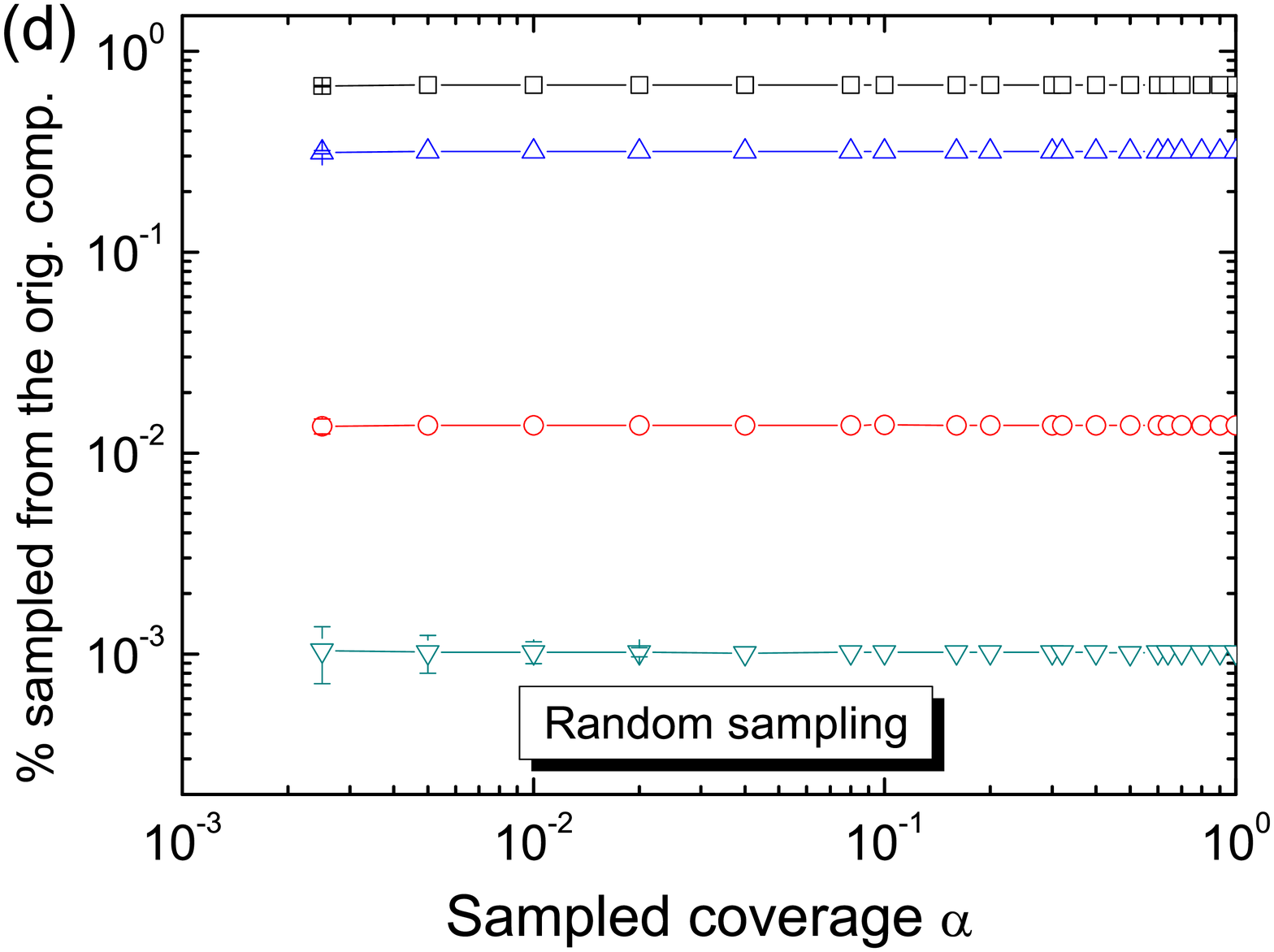}
\caption{ (Color online) For Wiki2007 data, component ratios in
the sampled networks (a), (b), and percentages sampled from the
components in the original network (c), (d). (a) It is surprising
that most of the network is a GSCC even at very low coverage,
indicating the importance of loops and clustering around high
in-degree nodes. The OUT component ratio slightly decreases as
$\alpha$ increases since the number of nodes in the GSCC increases
more quickly than the number of nodes in the OUT. (c) Conversely,
the percentage sampled from the OUT increases. As expected, the
other components shrink as a power of $1/N \approx \alpha^{-1}$.
(d) In the case of random sampling, the percentage sampled from
each component is almost constant. (b) On the other hand, the
component ratio changes substantially as the sampled coverage
increases. At small coverage, most of the nodes are disconnected,
but after a percolation threshold has been reached, the GSCC
emerges quickly, absorbing the other components. The labels for
(c), (d) are the same as those of (a), (b) and the numbers in
parentheses are the component ratio of the original network.}
\label{fig:Components}
\end{figure*}

\subsection{Components Ratios}

Here we focus both on the evolution of the bow-tie structure and
on the component from which the nodes are sampled -- noted in
Fig.~\ref{fig:Components}(c) and (d)  as ``\% sampled from the
orig. comp.''-- as the sampling coverage increases. BFS sampling
mainly covers the GSCC and OUT components, so the sizes of the IN
and TEND components in the sampled networks remain constant as
$\alpha$ increases. As coverage increases, the size ratio of the
GSCC -- the ratio of nodes in the current GSCC to the total number
of discovered nodes -- increases slightly as the GSCC absorbs
other components.

The main characteristics associated with random sampling are
described by percolation
phenomena~\cite{Albert2000,Cohen2000,Cohen2001}. When the sampling
coverage is small, most of the nodes are disconnected and belong
to the DISC and TEND components. As sampling coverage increases
past some percolation threshold, the GSCC emerges quickly and the
IN and OUT components form concurrently as shown in
Fig.~\ref{fig:Components}(b).

\section{Summary and Discussion}
\label{Summary_and_Discussion}
In summary, a comparison of BFS sampling to random sampling
indicates that differences in sampling method and coverage can
introduce biases that result in substantial mischaracterization of
the statistics of many structural properties in directed networks.
Moreover, the extent to which sampling biases will affect these
properties seems to depend heavily on the structure of the
original network. In comparing random sampling to BFS sampling on
seven different directed networks, including three versions of
Wikipedia, three different sources of sampled World Wide Web data,
and an Internet-based social network, we found that  differences
in sampling method and coverage affect both the bow-tie structure,
as well as the number and surface structure of strongly connected
components in sampled networks.  In addition, at low sampling
coverage (less than 40\%), the values of average degree, variance
of in- and out-degree, degree auto-correlation, and link
reciprocity in sampled networks are misestimated by at least 30\%,
and sometimes by as much as four orders of magnitude.  The
structural properties of BFS-sampled networks attain values within
10\% of the corresponding values in the original networks only
when sampling coverage is in excess of 65\%.

Most biases under random sampling seem to stem from the fact that
both out-degree and in-degree will be approximately equally
undersampled. This leads to underestimation of  average degree and
variances of in- and out-degree. At the same time, properties such
as reciprocity and auto-correlation are essentially constant
because of this equality in undersampling.  Biases under BFS
sampling arise from a confluence of factors:  by following only
outgoing links, BFS fails to cover the IN-component of directed
networks; BFS covers nodes of high in-degree at early times; the
core of BFS-sampled networks are tangled with many loops showing
high clustering; the in- and out-degrees of nodes at the growing
front are undersampled under BFS sampling. In combination, these
factors (and, possibly others) lead to overestimation of some
structural properties (average degree, variance in out-degree,
auto-correlation, and reciprocity) and underestimation of others
(variance in in-degree, number of SCCs, surface node ratios).  We
have demonstrated that for these reasons, if uniform random, or
BFS sampling is used to assemble a network, significant
corrections to degree, degree variance, auto-correlation,
reciprocity, some types of assortativity and component make-up
should be expected.

Though we have not examined it here, we suspect that there may be
an important interplay between sampling method, sampling coverage,
temporal changes, and sampled network topologies. The Wikipedia
data discussed earlier could be used to probe such effects, since
it captures snapshots of Wikipedia at different times during the
network's evolution. It would be interesting to quantify
differences in the effects (if any) of  BFS and random sampling on
time-varying or temporal networks~\cite{Holme2012}. A natural question, after analyzing the drawbacks of sampling procedures, will be how we can overcome such problems to get unbiased network samplings. A possible solution could be a combination of random and BFS samplings to get several unbiased structural properties. However, it is still challenging work to get unbiased samplings for every network properties. There are several papers suggesting unbiased sampling strategies for specific properties~\cite{Kurant2010,Kurant2011}.

The results presented in this paper have widespread implications
for conclusions that have been drawn regarding the structure (and
function) of some of the most ubiquitously studied real-world
networks, including the World Wide Web.  Since for many studied
real, directed networks only an incomplete link list is available,
either because the networks are too large to be fully recorded, or
because they change  too quickly to be captured by any sampling
procedure, our findings call into question the accuracy of
previous, reported results for the statistics of some of these
networks' structural properties. We may not know as much about the
structure of very large directed networks as has been supposed.

\begin{acknowledgments}
This work was partially supported by the research fund of Hanyang University (HY-2012-N) (S.-W.S.).
\end{acknowledgments}




\begin{thebibliography}{58}
\bibitem{note1}
The average is different from $\langle k \rangle'$ because the latter
is an average over random nodes, which in $\overline{j}_{x}$ the
nodes are chosen as end points of random links.
\end{thebibliography}
\end{document}